\documentclass[acmsmall]{acmart}
\AtBeginDocument{%
  \providecommand\BibTeX{{%
    Bib\TeX}}}
\usepackage[linesnumbered,ruled,vlined]{algorithm2e}
\usepackage{pifont}
\usepackage{svg} %
\usepackage{float}
\usepackage{graphicx} 
\usepackage{caption}
\usepackage{subcaption} 
\usepackage{enumitem}
\usepackage{tabularx}
\usepackage{amsmath}
\usepackage{booktabs}
\usepackage{caption}
\usepackage{multirow} 
\usepackage{array}
\usepackage{marginnote}
\usepackage{tcolorbox}
\usepackage{xcolor}

\newcommand{\kw}[1]{\text{{#1}}\xspace}
\newcommand{\hvq}{\kw{HVQ}}
\newcommand{\method}{\kw{DEG}}

\newcommand{\cheng}{}
\AtBeginDocument{%
  \providecommand\BibTeX{{%
    \normalfont B\kern-0.5em{\scshape i\kern-0.25em b}\kern-0.8em\TeX}}}

\newcolumntype{C}[1]{>{\centering\arraybackslash}m{#1}}

\newcommand{\revision}[1]{\textcolor{black}{#1}}
\newcommand{\ready}[1]{\textcolor{black}{#1}}

\setcopyright{rightsretained}
\acmYear{2025}

\acmJournal{PACMMOD}
\acmVolume{3}
\acmNumber{1 (SIGMOD)}
\acmArticle{29}
\acmMonth{2}
\acmPrice{}
\acmDOI{10.1145/3709679}

\begin{document}

\title{DEG: Efficient Hybrid Vector Search Using the Dynamic Edge Navigation Graph}

\author{Ziqi Yin}
\email{ziqi003@e.ntu.edu.sg}
\orcid{0009-0008-4221-2251}
\affiliation{
  \institution{Nanyang Technological University}
  \country{Singapore}
}

\author{Jianyang Gao}
\email{jianyang.gao@ntu.edu.sg}
\orcid{0009-0008-4684-3624}
\affiliation{
  \institution{Nanyang Technological University}
  \country{Singapore}
}

\author{Pasquale Balsebre}
\email{pasquale001@e.ntu.edu.sg}
\orcid{0009-0004-9454-2704}
\affiliation{
  \institution{Nanyang Technological University}
  \country{Singapore}
}

\author{Gao Cong}
\email{gaocong@ntu.edu.sg}
\orcid{0000-0002-4430-6373}
\affiliation{
  \institution{Nanyang Technological University}
  \country{Singapore}
}

\author{Cheng Long}
\email{c.long@ntu.edu.sg}
\orcid{0000-0001-6806-8405}
\affiliation{
  \institution{Nanyang Technological University}
  \country{Singapore}
}

\begin{abstract}
Bimodal data, such as image-text pairs, has become increasingly prevalent in the digital era. The Hybrid Vector Query (\hvq) is an effective approach for querying such data and has recently garnered considerable attention from researchers. It calculates similarity scores for objects represented by two vectors using a weighted sum of each individual vector's similarity, with a query-specific parameter $\alpha$ to determine the weight. Existing methods for \hvq typically construct Approximate Nearest Neighbors Search (ANNS) indexes with a fixed $\alpha$ value. This leads to significant performance degradation when the query's $\alpha$ dynamically changes based on the different scenarios and needs.

In this study, we introduce the Dynamic Edge Navigation Graph (\method), a graph-based ANNS index that maintains efficiency and accuracy with changing 
{\cheng $\alpha$ values}. It includes three novel components: (1) a greedy Pareto frontier search algorithm to compute a candidate 
neighbor set for each node, which comprises the node’s approximate nearest neighbors 
{\cheng for all possible $\alpha$ values;}
(2) a dynamic edge pruning strategy to determine the final edges from the candidate set and assign each edge an active range. This active range enables the dynamic use of the Relative Neighborhood Graph's pruning strategy based on the query's $\alpha$ values, skipping redundant edges at query time and achieving a better accuracy-efficiency trade-off; {\cheng and} (3) an edge seed method that accelerates the querying process. Extensive experiments on real-world datasets show that \method demonstrates superior performance compared to existing methods under varying 
{\cheng $\alpha$ values}.

\end{abstract}

\ccsdesc[500]{Information systems~Data management systems}
\ccsdesc[500]{Information systems~Proximity search}

\keywords{approximate nearest neighbor search, graph-based index, hybrid vector search}

\setcopyright{acmlicensed}
\acmJournal{PACMMOD}
\acmYear{2025} \acmVolume{3} \acmNumber{1 (SIGMOD)} \acmArticle{29} \acmMonth{2}\acmDOI{10.1145/3709679}

\maketitle

\section{Introduction}
\label{sec:intro}
Nearest Neighbor Search (NNS) in high-dimensional Euclidean space has been a fundamental component of many applications, including {\cheng databases~\cite{milvus2021, pan2024vector}}, information retrieval~\cite{ANNForIR, ANNForDPR}, data mining~\cite{ANNForDM}, recommendation systems~\cite{ANNForRS}, and generative artificial intelligence (GAI)~\cite{zhao2024retrievalaugmented}. However, due to the curse of dimensionality~\cite{curse1, curse2}, existing NNS methods~\cite{Jagadishidistance2005} often fail to meet practical efficiency requirements. To address this, Approximate Nearest Neighbor Search (ANNS) has been proposed, 
trading off some accuracy for significantly improved efficiency~\cite{li2020improving}.
%

Over the past decades, various ANNS techniques have been proposed~\cite{arora2018hdindex, BeygelzimerKL06Covertree, RamS19kdtree,wangSurveyLearningHash2018, hash2014survey, jegouProductQuantizationNearest2011, gongIterativeQuantizationProcrustean2013, DBLP:journals/pami/GeHK014,gao2024rabitq,wang2021comprehensive}, such as tree-based methods~\cite{arora2018hdindex, BeygelzimerKL06Covertree, RamS19kdtree}, hashing-based methods~\cite{wangSurveyLearningHash2018, hash2014survey}, quantization-based methods~\cite{jegouProductQuantizationNearest2011, gongIterativeQuantizationProcrustean2013, DBLP:journals/pami/GeHK014,gao2024rabitq}, and graph-based methods~\cite{wang2021comprehensive,fu2019fast,DBLP:journals/pami/FuWC22,DBLP:journals/is/MalkovPLK14,DBLP:conf/cvpr/HarwoodD16,malkovEfficientRobustApproximate2020}. 
Among them, graph-based methods provide superior performance compared to other methods~\cite{wang2021comprehensive,liApproximateNearestNeighbor2020}. 
These methods construct a graph 
to index the dataset
, where each node 
corresponds to an object in dataset 
and two nodes are associated with an edge if their distance satisfies some proximity property. 
At query time, a greedy search algorithm~\cite{fu2019fast,malkovEfficientRobustApproximate2020,wang2021comprehensive} is used to find approximate nearest neighbors. 
Most state-of-the-art graph-based ANNS indexes are constructed by approximating
the Relative Neighborhood Graph (RNG)~\cite{rng}. Its pruning strategy effectively removes some redundant edges based on edge length (distance between nodes), thereby achieving a better accuracy-efficiency tradeoff, as verified in~\cite{wang2021comprehensive} ({
details will be presented} in Section~\ref{sec:graph-anns}).

In recent years, researchers {
started to explore various} variants of ANNS queries~\cite{pan2024vector}, in order to handle more complex real-world scenarios. 
{
In particular, \cite{milvus2021} studies the multi-vector query, where each object is described by multiple vectors, and the similarity score is computed by aggregating 
each individual vector's similarity score.
} 
{For instance, intelligent video surveillance systems employ different vectors to represent the front face, side face, and posture of each individual recorded by the camera~\cite{baltruvsaitis2018multimodal}, which will subsequently be used for person identification.}


In this {
paper}, we investigate the Hybrid Vector Query (\hvq), which {
is one kind of} multi-vector query, {
where each object involves} two vectors. {
Specifically,} an \hvq aims to retrieve the approximate nearest neighbors according to the weighted sum of the distances {
based on the two} vectors, where {
the distance between 
vectors captures the dissimilarity between them and} a 
query-specific parameter $\alpha$ determines the weight of each vector's distance (as to be formulated in Section~\ref{sec:preliminary}). {
\hvq has many applications given the} ubiquity of bimodal data in real-life scenarios, such as image-text data~\cite{salvador2017learning,OpenImagesLocNarr,sharma2018conceptual}, spatio-textual data~\cite{chenLocationKeywordbasedQuerying2021}, and video-text data~\cite{miech19howto100m,baltruvsaitis2018multimodal}.
In the case of geo-textual objects, one feature vector is derived from the object's textual description using language processing techniques, such as BERT~\cite{devlin2018bert}, while the other feature vector represents its two-dimensional geographical coordinates. 
{\cheng A \hvq query can be issued} by a user {\cheng looking} for a `Japanese sushi restaurant' 
{\cheng near his/her} position, with the parameter $\alpha$ capturing the query’s preference over the trade-off between geographical proximity and semantic similarity. 
\revision{
For example, machine learning techniques are used to learn a query-dependent weight $\alpha$ for a query~\cite{liu2023}, thereby better capturing the query’s preference.} 
\revision{Another example is in the intelligent video surveillance scenario~ 
\cite{milvus2021}, where each person can be represented by a face vector and a posture vector. This approach can enhance the accuracy of person recognition, where the weight of each vector could be different depending on the quality of each vector, such as the resolution of the image.}

\noindent\textbf{{
Existing methods and limitations.}} 
{
Two methods~\cite{milvus2021} have been proposed for the \hvq problem.} 
{Specifically, {
the first one} constructs {
an ANNS index based on the weighted distance with $\alpha = 0.5$.}}
In the query phase, it {
simply uses this index (which is built with $\alpha = 0.5$) for handling queries (which can have arbitrary values of $\alpha$).}
{
The second one} constructs separate ANNS indexes, {
one} for each modality. During the query phase, it performs {\cheng search} on each index, retrieves similar objects in each modality, and then re-ranks {
all retrieved objects} to obtain the approximate nearest neighbors. 
{
Graph-based ANNS indexes can be 
integrated with these two methods given their superior performance over other {\cheng types of ANNS index}. 
}\revision{
Additionally, when one of the feature vectors represents geographical coordinates, \hvq 
can function as 
the semantic-aware spatial keyword query~\cite{qianSemanticawareTopkSpatial2018}. 
The previous work on semantic-aware spatial keyword query~\cite{chenS2RtreePivotbasedIndexing2020,qianSemanticawareTopkSpatial2018, DBLP:conf/edbt/TheodoropoulosN24} has encountered significant efficiency challenges, largely due to the curse of dimensionality of high-dimensional semantic vectors~\cite{curse1,curse2} (as to be detailed in Section~\ref{sec:related})}.

{
The two methods~\cite{milvus2021} 
both adopt the strategy of using a fixed}
$\alpha$ value (e.g., 0 or 0.5) during the index construction phase to build the index. \textbf{This is because existing 
ANNS indexes are constructed under the assumption that the distances between objects are certain and pre-determined
.} 
{For instance, graph-based ANNS indexes select neighbors for each node based on the edge lengths (the distances between objects), which assumes that the edge lengths are certain and pre-determined.}
{
Using a fixed $\alpha$ value} during the indexing phase {
would help} ensure that 
{the distances between objects} 
remain constant, 
{thereby maintaining the proximity property in the ANNS indexes for the given $\alpha$ value.} 
However, 
{
different queries can 
have different $\alpha$ values 
}
based on their needs and scenarios, the 
distances between objects 
would change 
as well, making the proximity property 
of the 
ANNS index ineffective and leading to severe performance degradation (as to be shown in Section~\ref{sec:motivations}).

\noindent\textbf{Challenges.} In this {
paper}, we aim to develop a graph-based ANNS index that is capable of maintaining high performance for \hvq with varying 
$\alpha$ values. 
The key lies in how to construct a graph-based ANNS index that is capable of handling varying $\alpha$ values, which presents three challenges: \textbf{(1) How to compute the candidate neighbor set for each node.} Existing graph-based ANNS indexes typically acquire hundreds of approximate nearest neighbors for each node as the candidate neighbor set, avoiding considering all nodes in the dataset and thereby improving construction efficiency. However, as $\alpha$ changes, the {\cheng distances between objects would change}, and the approximate nearest neighbors of each node may vary significantly. Therefore, it becomes challenging to compute a candidate set for each node that comprises the approximate nearest neighbors {\cheng for} varying $\alpha$ values. \textbf{(2) How to determine edges from the candidate set.} A key idea of existing state-of-the-art graph-based ANNS indexes is to use the pruning strategy of Relative Neighborhood Graph (RNG)
to prune some redundant candidate edges based on the edge lengths, thereby determining the final edges from the candidate set. According to the experimental evaluation~\cite{wang2021comprehensive}, this pruning strategy can significantly improve search performance. 
However, for the \hvq problem, the $\alpha$ value
varies at query time, making the edge lengths dynamic, and the RNG pruning strategy does not work. A straightforward solution is to fix an  $\alpha$ value to prune edges during the index construction phase. However, as the query $\alpha$ varies, the RNG's property becomes ineffective, thereby leading to significant performance degradation (as to be shown in Section~\ref{sec:motivations}). How to design a new edge pruning strategy that can maintain the RNG's property under varying $\alpha$ values is an open problem.
\textbf{(3)~How to select the seed for the graph index in the \hvq problem.}
According to~\cite{fu2019fast}, the start node (seed) of the graph impacts {
the search path length, which reflects} the search efficiency. To reduce the search path length, \cite{fu2019fast} proposes {
to use} the graph's approximate center as the seed. However, the center can vary significantly with different $\alpha$ values. 
An intuitive solution is to {
use} multiple approximate centers for different $\alpha$ values, but this {
would degrade the} search efficiency 
{\cheng since some of the start nodes may not be good one for a particular $\alpha$ value.}

\noindent\textbf{Our method.} 
In the first challenge, since $\alpha$ can be any arbitrary value in $[0,1]$, finding the nearest neighbor of a given node for each possible $\alpha$ value becomes unfeasible.
To address this, we propose to treat the problem as a multi-objective optimization problem, where the distance of each individual vector is considered as an objective function. Then, we 
use the Pareto frontier~\cite{ma2020efficient} 
as the candidate set, ensuring that the nearest neighbor is always within the Pareto frontier for varying $\alpha$. However, in our context, finding the exact Pareto frontier for each node is expensive. To address this, we propose the Greedy Pareto Frontier Search (GPS) algorithm. 
%
By iteratively exploring the neighbor of neighbor and searching for Pareto frontiers within the small set, GPS efficiently finds high-quality approximate Pareto frontiers.


To handle the second challenge, we propose a novel dynamic edge pruning strategy. This strategy aims to maintain the property of the RNG for pruning 
at varying $\alpha$ values, as this property is essential for enhancing the search performance~\cite{wang2021comprehensive}. 
To achieve this, 
we find that some edges {\cheng would} be pruned at certain $\alpha$ values based on the pruning strategy of RNG, while at other $\alpha$ values, they {\cheng would} be preserved. This motivates us to come up with the idea of assigning each edge {\cheng a range called \emph{active range}}, covering the 
$\alpha$ values for that edge, within which it {\cheng would} not be pruned by the RNG's property. During the query phase, the edge is activated only if the query's specific $\alpha$ falls within its active range; otherwise, the edge {\cheng would} be ignored. This strategy ensures that for each query $\alpha$ value, the graph can dynamically prune edges based on the RNG's property, so the remaining graph formed by the activated edges satisfies the RNG's property, thereby ensuring high performance.

To tackle the issue of choosing appropriate seeds for the graph index, we propose a new edge seed method. This method uses nodes that are farthest from the center under varying $\alpha$ values as seeds, i.e., edge seeds. Since edge seeds are far from each other, at query time, the greedy search will automatically start from the seed node closest to the query and ignore distant seeds, thus avoiding the efficiency problem caused by multiple start nodes.

Based on these proposed techniques, we develop a novel Dynamic Edge Navigation Graph (\method). During the index construction phase, \method constructs the index by 
inserting nodes one by one, similar to previous methods~\cite{malkovEfficientRobustApproximate2020,DBLP:journals/is/MalkovPLK14}. For each inserted node, \method performs the GPS algorithm over the partially built graph to obtain the candidate neighbor set. Then it applies the dynamic edge pruning strategy to determine the edges from the candidate set. Finally, it checks whether each newly inserted node can update the edge seed set. At query time, \method employs a variant of the greedy search algorithm, which dynamically skips some edges based on the query's $\alpha$ value and their active ranges.



The main contributions of this work are summarized as follows:

\begin{enumerate}[leftmargin=*,topsep=0pt]
\item We 
analyze 
the limitation of existing methods for the \hvq problem (Section~\ref{sec:motivations}). Specifically, these methods perform well for certain query $\alpha$ values but face significant performance degradation when the {\cheng $\alpha$ value} varies. 
This finding
has not been reported in previous literature.

\item We propose a new graph-based ANNS index called \method to maintain high performance across varying {\cheng $\alpha$ values}. It comprises three technical contributions: (i) a greedy Pareto frontier search algorithm to compute a candidate neighbor set for each node, comprising the node’s approximate nearest neighbors for varying $\alpha$ values; (ii) a dynamic edge pruning strategy that dynamically prunes edges at query time to maintain the property of the Relative Neighborhood Graph; and (iii) an 
edge seed method. To the best of our knowledge, this is the first ANNS index designed specifically for HVQ. 


\item We conduct extensive experiments on real-world datasets, which demonstrate that (i) \method demonstrates the best performance compared to all baselines across different $\alpha$ settings; {\cheng and} (ii) \method maintains high performance across different $\alpha$ settings without significant degradation. 

\end{enumerate}

\section{Related Work}
\label{sec:related}

\noindent\textbf{Approximate Nearest Neighbor Search.} To address the ANNS problem, various methods~\cite{LSHDatarIIM04,liuSKLSHEfficientIndex2014, DBLP:journals/pami/GeHK014, jegouProductQuantizationNearest2011, gongIterativeQuantizationProcrustean2013, ite_matsui_2018, fu2019fast, malkovEfficientRobustApproximate2020, wang2021comprehensive} have been proposed, which can be classified into four categories: tree-based methods~\cite{arora2018hdindex, BeygelzimerKL06Covertree, RamS19kdtree}, hashing-based methods~\cite{LSHDatarIIM04, wangSurveyLearningHash2018, hash2014survey, DBLP:journals/pami/HeoLHCY15, liuSKLSHEfficientIndex2014, DBLP:journals/vldb/ZhengZWNLJ22,DBLP:journals/tods/TaoYSK10, DBLP:journals/tkde/TianZZ24, DBLP:journals/pvldb/LuWWK20, DBLP:journals/pvldb/HuangFZFN15,DBLP:conf/compgeom/DatarIIM04,DBLP:conf/mm/TuncelFR02,DBLP:conf/sigmod/GanFFN12,DBLP:conf/sigmod/LeiHKT20, DBLP:conf/sigmod/LiYZXCLNC18, DBLP:journals/pvldb/GongWOX20}, quantization-based methods~\cite{jegouProductQuantizationNearest2011, gongIterativeQuantizationProcrustean2013, DBLP:journals/pami/GeHK014,ite_matsui_2018,gao2024rabitq}, and graph-based methods~\cite{wang2021comprehensive,fu2019fast,liApproximateNearestNeighbor2020,DBLP:journals/pami/FuWC22,DBLP:journals/is/MalkovPLK14,DBLP:conf/cvpr/HarwoodD16,malkovEfficientRobustApproximate2020,harwood2016fanng,chen2018sptag,iwasaki2015neighborhood,DBLP:journals/pami/FuWC22}. According to experimental evaluations~\cite{wang2021comprehensive,liApproximateNearestNeighbor2020}, graph-based methods demonstrate superior performance compared with other methods. 
This is because 
other methods typically attempt to partition vectors into buckets and route queries to a small number of close buckets for fast retrieval, which is challenging in high-dimensional space~\cite{curse1,curse2}. 
To the best of our knowledge, none of these ANNS techniques have been 
designed 
for the \hvq problem.


\noindent\textbf{Graph-based ANNS Indexes.} 
Most existing graph-based indexes~\cite{wang2021comprehensive,fu2019fast,liApproximateNearestNeighbor2020,DBLP:journals/pami/FuWC22,DBLP:journals/is/MalkovPLK14,DBLP:conf/cvpr/HarwoodD16,malkovEfficientRobustApproximate2020,harwood2016fanng,chen2018sptag,iwasaki2015neighborhood,DBLP:journals/pami/FuWC22} are derived from four fundamental types of graphs: Delaunay Graph (DG)~\cite{dgraph}, Relative Neighborhood Graph (RNG)~\cite{rng}, K-Nearest Neighbor Graph (KNNG)~\cite{knngraph}, and Minimum Spanning Tree (MST)~\cite{mst}. 
According to the experimental evaluation~\cite{wang2021comprehensive}, RNG-based ANNS indexes deliver state-of-the-art performance 
due to its pruning strategy. 


\noindent\textbf{Variants of ANNS Query.} To address more complex real-world scenarios, several variants of the ANNS query have been proposed~\cite{pan2024vector}. These variants of ANNS queries have been identified as a promising future research direction in recent vector database studies~\cite{milvus2021,guo2022manu}. For instance, \cite{wei2020analyticdbv} \revision{has introduced a 
hybrid ANNS query} with attribute filtering to retrieve similar vectors that satisfy boolean predicates over their attributes. \cite{milvus2021} introduces the multi-vector query, from which we define HVQ, and develops two solutions (as detailed in Section~\ref{sec:preliminary-motivations}). 
However, both of these solutions have limitations in handling queries with different $\alpha$ (as shown in Section~\ref{sec:preliminary-motivations}). 
To the best of our knowledge, none of the existing studies have identified or attempted to address these limitations, and answering multi-vector query is an open problem.


\noindent\revision{\textbf{Hybrid Search.} HVQ is 
a variant of hybrid search, which has been researched extensively. Existing studies 
aim to efficiently identify  top-$k$ objects ranked by monotone aggregation functions. 
The Threshold Algorithm (TA) and the No Random Access (NRA) algorithm~\cite{fagin2001optimal} use pre-sorted lists for each attribute, but they face challenges with multi-dimensional data, 
as pre-computing distances and generating sorted lists is computationally expensive. To overcome this limitation, RR$^*$-tree~\cite{franzke2016indexing} uses reference objects to generate low-dimensional embeddings for each data object, which are then indexed by R-tree for efficient query processing.
However, these approaches are limited to low-dimensional spaces and degrade to no better than a linear scan in high-dimensional settings. 
}

\revision{There exist some attempts that investigate the integration of existing hybrid search techniques with hybrid vector search. Specifically, \cite{budikova2013towards} develops a distributed system based on the \textsf{Fusion} baseline, employing a modified TA algorithm to filter vectors fetched from distributed machines. 
\cite{milvus2021} uses the NRA algorithm
to determine when the \textsf{Fusion} baseline should stop increasing $k'$ in order to obtain exact top-$k$ results. However, the experimental results~\cite{milvus2021} show that integrating  NRA  with the \textsf{Fusion} baseline has extremely high computational cost, 
and iteratively increasing the $k'$ in the \textsf{Fusion} baseline is the most effective strategy.
}

\revision{
There also exists recent work on exploring hybrid vector search for special cases, such as semantic-aware spatial keyword queries~\cite{chenS2RtreePivotbasedIndexing2020,qianSemanticawareTopkSpatial2018, DBLP:conf/edbt/TheodoropoulosN24}, where one attribute is a geographical coordinate and the other is a high-dimensional embedding. 
It takes geo-location and 
embedding as inputs to find top-$k$ objects based on a weighted sum of geographical and embedding distances. However, these methods often focus on exact top-$k$ results, and it is still an open problem to design efficient querying algorithms. 
Details on these methods are provided in the appendix~\footnote{\url{https://github.com/Heisenberg-Yin/DEG/blob/main/SIGMOD2024_DEG_ready.pdf}} due to the page limitation.}


\section{Preliminary and MOTIVATIONS}
\label{sec:preliminary-motivations}

\subsection{Problem Statement}
\label{sec:preliminary}
We 
proceed to define the Hybrid Vector Query (\hvq).

\noindent\textbf{Dataset.} We consider a dataset $D$ consisting of $N$ objects. Each object $o\in D$ is characterized by two features: (1) {\cheng one} feature vector, denoted by $o.e$,  defined in a $d$-dimensional Euclidean space $E^{d}$, where $d$ is typically hundreds, and (2) {\cheng the other} feature vector, denoted {\cheng by} $o.s$, defined in an $m$-dimensional Euclidean space $E^{m}$, where $m$ 
may vary from two to hundreds, depending on applications.

\noindent\textbf{Hybrid Vector Query (\hvq):} Given a dataset $D$, a hybrid vector query $q=\langle e, s, \alpha, k\rangle$ consists of 
{\cheng two feature vectors $q.e \in E^{d}$ and $q.s \in E^{m}$,}
a hyperparameter $q.\alpha\in [0,1]$, and the number of objects to be returned $q.k$. 
\hvq aims to retrieve $k$ objects with the {\cheng minimum} 
distance $Dist(q,o)$:
\begin{equation}
\begin{aligned}
    Dist(q, o) = q.\alpha \times \delta_e(q, o) + (1-q.\alpha) \times \delta_s(q, o),
\end{aligned}
\label{eq:hybird_distance}
\end{equation}
where $\delta_e(q, o)=\frac{\delta(q.e, \ o.e)}{e_{max}}$ and $\delta_s(q, o)=\frac{\delta(q.s,\ o.s)}{s_{max}}$. Here, $\delta(x, y)$ represents the Euclidean distance between vectors $x$ and $y$, and $e_{max}$ (resp. $s_{max}$) denotes the maximum Euclidean distance between $o.e$ (resp. $o.s$) of any two objects in the dataset and is used as a normalization factor. 
Following 
previous work~\cite{milvus2021}, we use the weighted aggregated score as the similarity metric, and $q.\alpha \in [0, 1]$ is a 
parameter that allows to set preferences between the two vectors at query time. 

\noindent\textbf{Problem Statement.} In this study, we aim to develop a graph-based ANNS index for the \hvq problem that maintains both accuracy and efficiency under 
different query $\alpha$ values.

\noindent\revision{
\textbf{Comparison of HVQ and Hybrid Queries.} Several types of hybrid queries are related to \hvq, and we next discuss their differences. 
Early studies on hybrid queries~\cite{fagin2001optimal,franzke2016indexing} typically assume that each object has multiple attributes, such as price, and aim to identify the top-k objects ranked by monotone aggregation functions, e.g., the sum of these attributes. These studies 
focus on numeric or low-dimensional attributes (e.g., geo-location) and aim to find the exact top-k results (as to be detailed in Section~\ref{sec:related}). Due to the curse of dimensionality~\cite{curse1,curse2}, they are not suitable for the \hvq problem. 
When one of the vectors is low-dimensional coordinates, \hvq can function as semantic-aware spatial keyword queries~\cite{qianSemanticawareTopkSpatial2018}, which differ from traditional spatial keyword queries. However, existing methods for semantic-aware spatial keyword queries~\cite{chenS2RtreePivotbasedIndexing2020,qianSemanticawareTopkSpatial2018, DBLP:conf/edbt/TheodoropoulosN24} often suffer from severe efficiency challenges (as to be detailed in Section~\ref{sec:related}).}

\subsection{Graph-based ANNS methods}
\label{sec:graph-anns}

\begin{figure}[!t]
\begin{center}
\subcaptionbox{HNSW.\label{fig:hnsw}}{
\includegraphics[width=0.35\columnwidth]{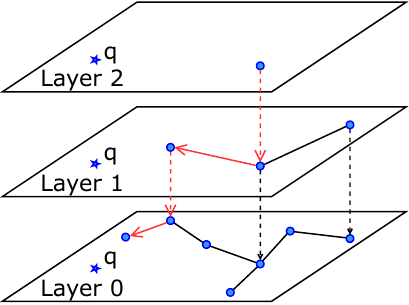}
}
\subcaptionbox{Pruning Strategy of RNG.\label{fig:rng}}{
\includegraphics[width=0.35\columnwidth]{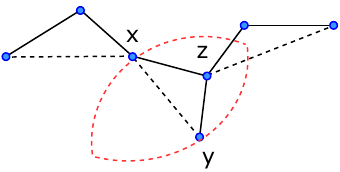}
}
\caption{Figure~\ref{fig:hnsw} illustrates the HNSW index. Figure~\ref{fig:rng}~illustrates the pruning strategy of the Relative Neighborhood Graph (RNG).}
\label{fig:hnsw-rng}
\end{center}
\end{figure}

According to experimental evaluations~\cite{liApproximateNearestNeighbor2020, wang2021comprehensive}, graph-based approaches have demonstrated superior performance compared to other ANNS techniques. 
They build a graph $G(V, E)$ where each node $x \in V$ corresponds to an object $o \in D$. The Hierarchical Navigable Small World graph (HNSW)~\cite{malkovEfficientRobustApproximate2020}, as illustrated in Figure~\ref{fig:hnsw}, is a 
state-of-the-art method. 
It comprises several layers, where layer 0 contains all objects, and each upper layer $i\ge1$, randomly retains a subset of the objects from layer $i-1$. Within each layer, a vertex is connected to several approximate nearest neighbors, while across adjacent layers, two vertices are connected only if they represent the same vector data.

At 
query time, the HNSW algorithm starts the search from the vertex in the top layer, as shown in Figure~\ref{fig:hnsw}. 
{At each layer, the HNSW algorithm accesses all neighbors of the current vertex and checks whether there exists a neighbor closer to the query than the current vertex. If so, it moves to the neighbor nearest to the query. This process continues within the layer until no closer neighbor to the query can be found.
It then proceeds to the next layer and repeats the same procedure, using the current vertex as the starting vertex. This process continues until reaching layer 0.} 
At layer 0, it conducts a greedy search algorithm, a method commonly used by 
graph-based ANNS indexes~\cite{wang2021comprehensive,fu2019fast,liApproximateNearestNeighbor2020,DBLP:journals/pami/FuWC22,DBLP:journals/is/MalkovPLK14,DBLP:conf/cvpr/HarwoodD16}. Specifically, it maintains two sets, a candidate set $\mathcal{S}$ (a min-heap) and a result set $\mathcal{R}$ (a max-heap), where $R$ stores the approximate nearest neighbors that have currently been found for the query, and $S$ stores candidates that can potentially improve $R$. At each iteration, the object with the smallest distance in $\mathcal{S}$ is popped, and its neighbors are evaluated. If a neighbor has not been visited before and its distance from the query is smaller than the maximum distance in $\mathcal{R}$, the neighbor is added to $\mathcal{S}$ and $\mathcal{R}$. Here, new elements are inserted into $\mathcal{R}$ continuously.
To avoid the high maintenance cost when $\mathcal{R}$ becomes large, which would reduce search efficiency, a hyperparameter $ef_{search}$ is used to control its size. If the size of $\mathcal{R}$ exceeds $ef_{search}$, the object with the maximum distance in $\mathcal{R}$ is removed. The candidate set $\mathcal{S}$ is unbounded in size, as empirically it does not significantly reduce efficiency. The search stops and returns the $k$ nearest objects in $\mathcal{R}$ when the minimum distance in $\mathcal{S}$ is larger than the maximum distance in $\mathcal{R}$. The hyperparameter $ef_{search}$ controls the accuracy-efficiency trade-off at query time.


Clearly, the graph structure plays a key role in graph-based ANNS indexes, as verified in the work~\cite{wang2021comprehensive}. According to~\cite{wang2021comprehensive}, 
state-of-the-art graph-based ANNS indexes (e.g., HNSW) are mostly constructed by approximating the Relative Neighborhood Graph (RNG)~\cite{rng}
, and we next explain the RNG. 

\noindent\textbf{Relative Neighborhood Graph (RNG).} The RNG $G(V, E)$ constructed on a dataset $D$ has the following property: For $x, y \in V$, if $x$ and $y$ are connected by an edge $e \in E$, then for $\forall z \in V$, either $\delta(x, y) < \delta(x, z)$ or $\delta(x, y) < \delta(y, z)$. In other words, the longest edge of a triangle in RNG will be pruned. \revision{
Note that the RNG property applies only to metric distances, as non-metric or non-linear distances do not satisfy the triangle inequality.}

Figure~\ref{fig:rng} illustrates this property of RNG. The dashed edge $(x, y)$ represents a potential edge and is pruned from RNG because it is the longest edge in the triangle $(x, y, z)$.
This pruning strategy removes some redundant neighbors and makes the neighbors distribute omnidirectionally, thereby reducing 
redundant search on the ANNS index~\cite{wang2021comprehensive}. For example, in the triangle $(x, y, z)$, if all three edges are preserved, there exist two paths from $x$ to $z$: $\{(x, z)\}$ and $\{(x, y), (y, z)\}$, which result in redundant calculations. According to the experimental evaluation~\cite{wang2021comprehensive}, the pruning strategy of RNG can improve search performance significantly. 


However, the time complexity of constructing the exact RNG on dataset $D$ is $O(N^3)$~\cite{jaromczyk1991constructing}. Therefore, many techniques~\cite{fu2019fast,malkovEfficientRobustApproximate2020,harwood2016fanng,chen2018sptag,iwasaki2015neighborhood,DBLP:journals/pami/FuWC22,liApproximateNearestNeighbor2020} have been developed to approximate RNG and construct an ANNS index. Take HNSW as an example, which constructs 
RNG by continuously inserting nodes. For each node, the key is how to determine edges for each node. The neighbor selection strategy in HNSW consists of two steps, which is also widely used by other graph-based ANNS indexes~\cite{fu2019fast,malkovEfficientRobustApproximate2020,harwood2016fanng,iwasaki2015neighborhood, DBLP:journals/is/MalkovPLK14}: (1) Obtaining $ef_{construction}$ approximate nearest neighbors as the candidate neighbor set, which avoids using all nodes in the graph as candidates, thus improving index construction efficiency. To obtain the candidate set, it performs a greedy search over the partially built graph; (2) Using the RNG's pruning strategy on the candidate set to prune some redundant candidate set and obtain the final $M$ neighbors, thereby achieving a better accuracy-efficiency trade-off. 
Here, $ef_{construction}$ is a hyperparameter that controls the candidate set size, and $M$ is a hyperparameter that determines the maximum number of neighbors per node. 

\subsection{Motivations}
\label{sec:motivations}
\noindent\textbf{Baselines.} To the best of our knowledge, very little work has been done for the \hvq problem and previous work only presents two simple solutions by adapting existing ANNS indexes~\cite{milvus2021}, which are detailed below:

\begin{itemize}[leftmargin=*, topsep=0pt]
    \item \textsf{Fusion (abbr. F)}: This method builds an ANNS index based on the hybrid distance by fixing $\alpha = 0.5$ in Equation~\ref{eq:hybird_distance}. During the query phase, it searches the built 
    ANNS index 
    to obtain the approximate result. The accuracy-efficiency trade-off is controlled by the search algorithm's hyperparameter (e.g., $ef_{search}$).
    \item \textsf{Merging (abbr. M)}: This method builds an index for each
    %
    feature vector separately. During the query 
    phase, it issues a top-$k'$ query for each query feature 
    on the corresponding index, where $k' \geq k$. All the returned 
    objects from every query feature are then reranked based on Equation~\ref{eq:hybird_distance} to find the approximate top-$k$ results. The accuracy-efficiency trade-off is largely affected by the hyperparameter $k'$.
\end{itemize}
We integrate the \textsf{Fusion} method with the state-of-the-art ANNS index, \textsf{HNSW}~\cite{malkovEfficientRobustApproximate2020}. 
This method is denoted as \textsf{HNSW}$_{\textsf{F}}$. For the \textsf{Merging} method, we use \textsf{HNSW} to build an ANNS index for each modality separately. When one of the feature vector's dimensions is low, 
we use the R-Tree~\cite{beckmann1990r} to index this modality and issue an exact top-$k'$ query instead 
as the R-tree will perform better. This method is denoted as \textsf{HNSW}$_{\textsf{M}}$.

\begin{figure}[!t]
\centering
\includegraphics[width=0.7\columnwidth]{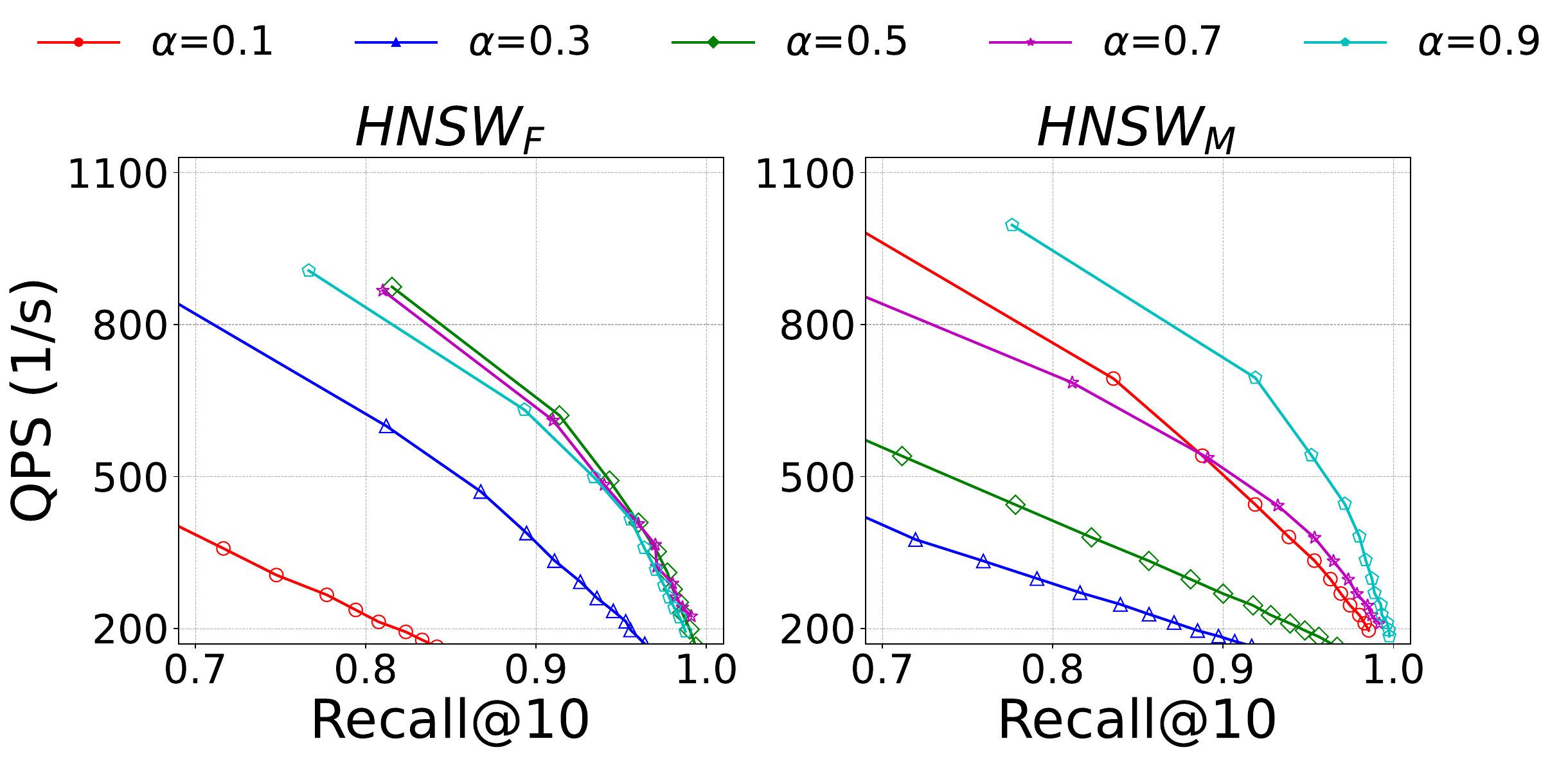}
\vspace*{-1em}
\caption{The experiment results of \textsf{HNSW}$_{\textsf{M}}$ and 
\textsf{HNSW}$_{\textsf{F}}$ on the OpenImage dataset (up and right is better).}
\label{fig:exp-motivation}
\vspace*{-1em}
\end{figure}

\noindent\textbf{Limitations of Baselines.} Both baselines share a common idea: fixing the $\alpha$ value (e.g., 0, 0.5, 1)
to build the index
during the 
construction phase. \textbf{This is because existing ANNS indexes are 
designed for the setting 
in which
distances between objects 
are certain and pre-determined.} Take graph-based ANNS indexes as an example, they typically 
selects approximate nearest neighbors 
for each node as the edge candidate set and pruning candidate edges based on edge length (distance between nodes). However, for the \hvq problem, as the query's $\alpha$ changes, the distances between objects may also change. The 
dynamic nature of
distances in the \hvq problem 
exceeds
the capabilities of existing ANNS techniques to handle
, and a straightforward approach is to fix the value of $\alpha$ during the indexing phase.

\textbf{The limitation of fixing $\alpha$ values to build ANNS indexes in the \hvq problem is that while this approach performs well for some query $\alpha$ values, it faces significant performance degradation for others.} For example, 
\textsf{HNSW}$_{\textsf{F}}$ 
constructs the index by fixing $\alpha$ at 0.5, ensuring that the edges satisfy the RNG's properties when $\alpha$ is 0.5. However, during the query phase, when $q.\alpha$ deviates
significantly from 0.5, the properties of RNG quickly
become ineffective, leading to a significant degradation in performance (as to be shown later). 
\textsf{HNSW}$_{\textsf{M}}$ 
constructs separate indexes, 
each of which considers only one feature vector while neglecting the other. Therefore, each index can only retrieve similar objects within a modality. When $q.\alpha$ is around 0.5 during the query time, the separate indexes struggle to fetch high-quality candidates that are similar in both modalities for reranking, resulting in severe performance degradation (as to be shown later).

To illustrate the baselines' limitations, We evaluate \textsf{HNSW}$_{\textsf{F}}$ and \textsf{HNSW}${_\textsf{M}}$ on the \textsf{OpenImage} dataset, which contains 500K images and 
textual descriptions. The textual content and images are transformed into embeddings using BERT~\cite{devlin2018bert} and ViT~\cite{dosovitskiy2020image}, respectively. The efficiency-accuracy trade-off results are shown in Figure~\ref{fig:exp-motivation}
(hyperparameters details can be found in Section~\ref{sec:exp-setup}).
We make the following observations: 
(1) \textsf{HNSW}$_{\textsf{F}}$ 
performs the best when $q.\alpha$ is 
(e.g., 0.5) 
and degrades significantly when $q.\alpha$ deviates from 0.5 (e.g., 0.1 and 0.9);
(2) \textsf{HNSW}$_{\textsf{M}}$ 
performs better when $q.\alpha$ is close to 0 and 1 (e.g., 0.1 and 0.9) and degrades significantly when $q.\alpha$ is close to 0.5. 
This validates our analysis that adopting existing graph-based ANNS methods as baselines can perform well only for certain values of $q.\alpha$, but they exhibit significant performance degradation under different $q.\alpha$ values. These limitations motivate us to develop a new graph-based ANNS index that maintains high performance across varying $\alpha$ values.

\begin{figure*}[!t]
\begin{center}
\includegraphics[width=\textwidth]{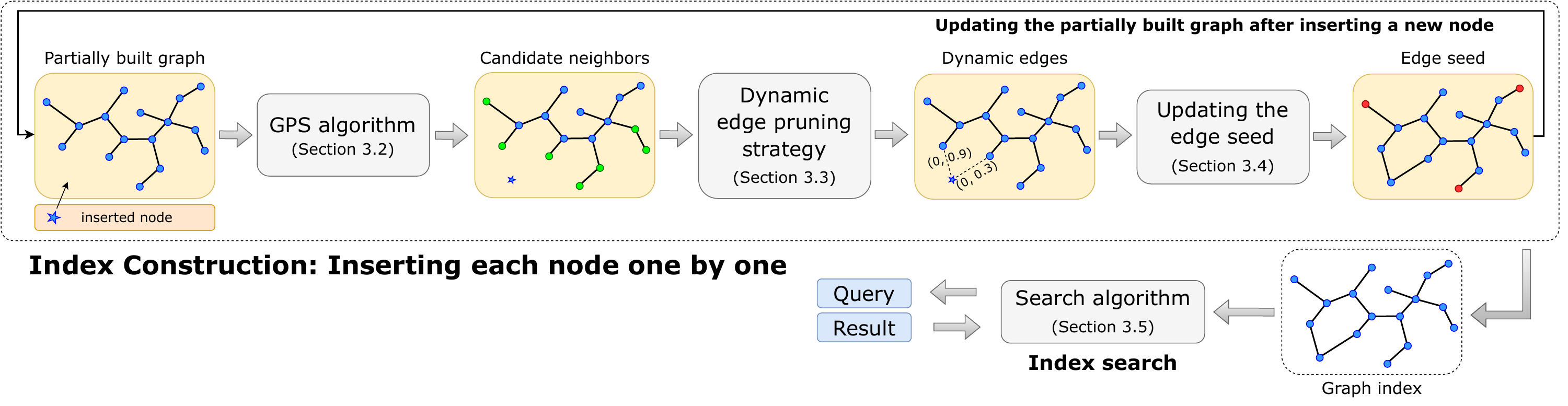}
\caption{The framework of the \method, including the index construction phase and index search phase.}
\label{fig:framework}
\end{center}
\end{figure*}

\section{The \method Method}
\label{sec:method}
\subsection{Overview}
\label{src:overview}

To overcome the limitations of baselines, and address the three challenges discussed in Section~\ref{sec:intro}, we develop a new graph-based ANNS index, namely Dynamic Edge Navigation Graph (\method), which includes three components, each addressing a challenge. We proceed to give an overview of the three components. 



To tackle the first challenge of computing a candidate neighbor set for each node, 
we propose 
the GPS algorithm. The high-level idea of the GPS algorithm can be summarized in two points: 1) Finding the nearest neighbor of a node at any $\alpha$ value is equivalent to solving the problem of minimizing the distance function (Equation~\ref{eq:hybird_distance}) at that $\alpha$ value. Finding solutions for each possible $\alpha \in [0, 1]$ is unfeasible. Then, we treat it as a multi-objective optimization problem, with each vector distance as an objective. 
To solve this, we 
compute the Pareto frontier as the solution, i.e., the neighbor candidate set, ensuring the nearest neighbor for any $\alpha$ value is included; 2) Finding the exact Pareto frontier is expensive.
Therefore, the GPS algorithm aims to find approximate Pareto frontiers to reduce index construction costs. The core idea of the GPS algorithm is that a neighbor’s neighbor is likely to be a neighbor. By iteratively exploring neighbors of neighbors, it continuously optimizes the approximate Pareto frontiers.

To address the second challenge of determining edges from the candidate set, we propose a dynamic edge pruning strategy that utilizes the RNG's pruning strategy to dynamically prune edges in \method.
Our key idea is as follows: 1) We assign an active range to each edge, covering the suitable $\alpha$ values for that edge, within which that edge will not be pruned by the RNG's property; and 2) at query time, we use an edge for routing only if the query's $\alpha$ value intersects with its active range; otherwise, we ignore it. To achieve this, we take $\alpha$ into account in the pruning strategy of RNG and compute an active range for each edge such that for any $\alpha$ value, the remaining graph formed by the activated edges satisfies the RNG's property, thereby ensuring high performance. 

To address the third challenge of finding the appropriate seed for the graph index, we propose a new edge seed method. This method uses nodes that are farthest from the center for varying $\alpha$ values as seed, i.e., edge seed. Compared to using multiple approximate centroids, edge seed ensures that they are far from each other. In greedy search, only edge nodes close to the query are activated, while others are ignored due to their large distance, thus avoiding the efficiency issue caused by multiple start nodes.

Figure~\ref{fig:framework} illustrates the framework of \method, comprising the index construction phase and the index search phase. In the index construction phase, \method builds the graph index by continuously inserting new nodes, similar to HNSW. It performs the GPS algorithm over the partially built graph to obtain candidate neighbors of the inserted node (shown as green nodes). Then, it utilizes the dynamic pruning strategy to prune some candidate edges, deriving the final dynamic edges (shown as dashed lines), where each edge is assigned an active range. Finally, it updates the edge seed (shown as red nodes). In the index search phase, the edge seed is used as the starting node, and a variant of the greedy search algorithm is employed, which dynamically skips some edges based on their active ranges and the query's $\alpha$ value. 

In the rest of this section, we first present the three main components of \method, namely (1) a greedy Pareto frontier search algorithm, called the GPS algorithm (Section~\ref{sec:candidate-acquisition}); (2) a dynamic edge pruning strategy (Section~\ref{sec:D-RNG}); and (3) an edge seed method (Section~\ref{sec:construction}). Then we present the search algorithm in Section~\ref{sec:search}.

\vspace*{-0.5em}
\subsection{Candidate Neighbor Acquisition}
\label{sec:candidate-acquisition}
As previously discussed, finding the nearest neighbor of a node at any $\alpha$ value is equivalent to solving a multi-objective optimization problem.
We formalize this problem as follows: Given an object $p \in D$, for other objects $x \in D\setminus \{p\}$, the multi-objective function is defined as $f(p, x): x \rightarrow \mathbb{R}^2$, where $f_1(p, x) = \delta_e(p, x)$ and $f_2(p, x) = \delta_s(p, x)$. 
We propose using the Pareto frontier as the solution to this problem. Next, we introduce the concept of the Pareto frontier and explain why it is used as the 
candidate set.



\noindent\textbf{Pareto Frontier~\cite{ma2020efficient}.} Consider a dataset $D$ and a multi-objective optimization problem described by $f(x)\colon x \rightarrow \mathbb{R}^c$, where $x \in D$. Each function $f_i(x) \rightarrow \mathbb{R}$ represents the objective function of the $i$-th task to be minimized, with $i\in C$ and $C = \{1, 2, \cdots, c\}$. For any $x, y \in D$, $x$ dominates $y$ if and only if (1) $\forall i \in C$, $f_i(x) \leq f_i(y)$ and (2) $\exists j \in C$, $f_j(x) < f_j(y)$. An object $x \in D$ is considered Pareto optimal if no other object in $D$ dominates $x$. The Pareto frontier, also called the Pareto set, comprises all Pareto optimal objects in $D$.

\begin{figure}[!t]
\begin{center}
\subcaptionbox{Pareto Frontier.\label{fig:pareto}}{
\includegraphics[width=0.35\columnwidth]{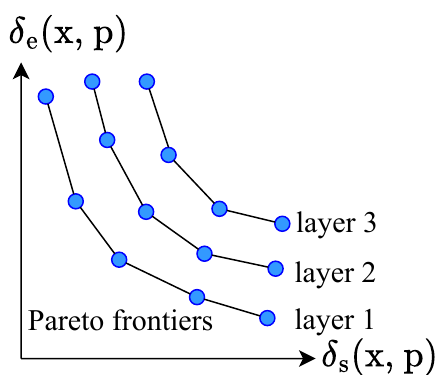}
}
\subcaptionbox{Edge seed method.\label{fig:seed}}{
\includegraphics[width=0.35\columnwidth]{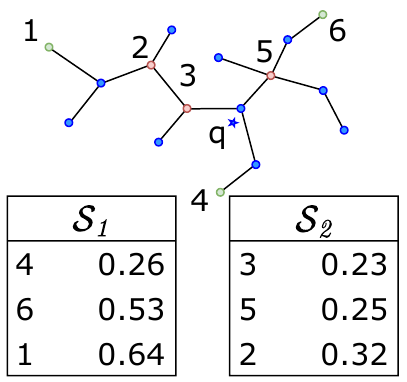}
}
\caption{Figure~\ref{fig:pareto}~illustrates the Pareto frontiers and Figure~\ref{fig:seed}~illustrates the difference between the edge seed acquisition method and the multiple centroids method.}
\end{center}
\end{figure}

Figure~\ref{fig:pareto} illustrates the Pareto frontier. For a node $p$, $\delta_s(x, p)$ and $\delta_e(x, p)$ denote the distance of each individual vector. 
The nodes within layer 1 lie at the bottom of the graph and constitute the Pareto frontier described above. Nodes in other layers will also be collected into the candidate neighbor set, as will be explained later. 
Next, we present the theorem that explains why the Pareto frontier can be used as the candidate set.

\begin{theorem}
\label{theorem:nearest}
We denote the Pareto Frontier of the multi-objective function $f(p, x)$ as $PF(D, p) \subset D\setminus\{p\}$. For any $\alpha \in [0,1]$, 
the nearest neighbor of $p$ is contained in $PF(D, p)$. 
\end{theorem}
\vspace*{-0.5em}
Due to page limitations, the proof is provided in the appendix. Therefore, we can compute $PF(D, p)$ for each object $p \in D$ as the 
candidate set, ensuring the nearest neighbors are always within $PF(D, p)$ when $\alpha$ varies. However, in a dataset with millions of objects, $PF(D, p)$ may contain only a dozen objects (e.g., 10), while we usually need hundreds of objects (e.g., 200) as the edge candidates to select the final edges. To address this, we can choose multiple layers of $PF(D, p)$. Specifically, after finding the $PF(D, p)$, we remove the objects within $PF(D, p)$ from the dataset and search for a new $PF(D, p)$ until we obtain enough edge candidates. 
Figure~\ref{fig:pareto} illustrates this, where layers 2 and 3 will also be collected as part of the candidate set.

The positive news is that for a given node $p$, with the distances $\delta_e(p, x)$ and $\delta_s(p, x)$ from other nodes to $p$ already calculated, finding $PF(D, p)$ is equivalent to 
finding the two-dimensional skyline~\cite{borzsony2001skyline}, which has efficient solutions~\cite{khalefa2008skyline,zhiyonghuang2006continuous,papadias2005progressive,kalyvas2017survey}. 
However, these methods cannot be efficiently applied to our problem directly. 



\noindent\textbf{Greedy Pareto Frontier Search.} 
To address this, we introduce a novel Greedy Pareto Frontier Search algorithm (called GPS algorithm) for searching approximate Pareto frontiers on a partially built graph. The core idea of the GPS algorithm is that the neighbor of a neighbor is more likely to be a neighbor. It continuously explores the neighbors of neighbors and searches for the multi-layer Pareto frontier (skylines) within this small set. This approach allows us to obtain a high-quality approximate Pareto frontier and improve efficiency.
\begin{algorithm}[t]
    \caption{\revision{GPS($G$, $q$, $ep$, $ef_{constrution}$)}}
    \small
    \label{alg:pareto-frontier-search}
    \KwIn{A partially built graph $G$, a query $q$, start node set $ep$, candidate pool size $ef_{construction}$}
    \KwOut{Multi-layer Pareto frontiers $\{PF^{1}(q), \cdots, PF^{l}(q)\}$ for $q$, where 
    $\sum_{i = 1}^l |PF^{i}(q)| \leq ef_{construction}$}
    
    \BlankLine
    
    \SetKwFunction{GPS}{\textsc{GPS}}
    \SetKwProg{Fn}{Procedure}{:}{}
    \Fn{\GPS{$G, q, ep, ef_{constrution}$}}{
    
        $Res \leftarrow \emptyset$ \tcp*{\textsf{set of the pareto frontiers}}
        $Vis \leftarrow \emptyset$ \tcp*{\textsf{set of the visited nodes}}
        $Flag \leftarrow \emptyset$ \tcp*{\textsf{set of the explored nodes}}    
        \ForEach{$v \in ep$}{
            $Res.\text{add}(v)$\; 
            $Vis.\text{add}(v)$\;
        }
        $Res \leftarrow FindPF(Res, ef_{constrution})$\;  

    \While{$|Res| < ef_{construction}$}{        
            $NNS \leftarrow \emptyset$ \tcp*[r]{\textsf{set of unexplored new nodes}} 
    
            \ForEach{$S_k \in Res$} 
            {
                $NNS \leftarrow \{v \in \mathcal{S}_k \mid  v \notin Flag\}$\;                 
                \If{$NNS \neq \emptyset$}{
                    \textbf{break}\;        
                }
            }
            \If{$NNS = \emptyset$} 
            {
                \textbf{break}\;  
            }
            
    
            \ForEach{$u \in NNS$}{
                $Flag.\text{add}(u)$  \tcp*{\textsf{mark $u$ as old node}}
                \ForEach{$v \in neighbour(u)$}{
                    \If{$v \notin vis$}{
                        $Res.\text{add}(v)$\;
                        $Vis.\text{add}(v)$\;
                    }
                }
            }
            \tcp{\textsf{update the pareto frontiers using existing algorithm}} 
            $Res \leftarrow FindPF(Res, ef_{constrution})$\;
        }
        \Return $Res$\;
    }
\end{algorithm}

The detailed procedures of the GPS algorithm are summarized at Algorithm~\ref{alg:pareto-frontier-search}. 
It maintains a candidate set $Res$ to store the \revision{discovered approximate} Pareto frontiers, an indicator $Vis$ to mark whether a node was added to $Res$, and another indicator $Flag$ to mark whether a node's neighbors have been evaluated and added to $Res$ (lines 2-4).  \revision{If a node's neighbor is not evaluated,} we call it a new node in $Res$. The GPS algorithm first adds all the start nodes (the seed) into $Res$ as the  \revision{initial candidates} 
and marks them as visited (lines 5-7).  \revision{Then it organizes these initial candidates into multi-layer Pareto frontiers using a modification of existing algorithm~\cite{borzsony2001skyline}, referred to as $FindPF$ (line 8). Specifically, $FindPF$ repeatedly finds the Pareto frontier (skyline) within the candidate set $Res$ using the existing algorithm~\cite{borzsony2001skyline}, adds it to the result set, and removes it from the candidate set. When the result set reaches the size bound, it returns the first $l$ layer Pareto frontiers that satisfy the size limit.
Due to page limitations, the details of the algorithm are provided in the appendix.}
\revision{Next}, the GPS algorithm iteratively performs a greedy search over graph $G$ to optimize the discovered approximate 
Pareto frontiers (lines 9-23). Specifically, it scans each layer of the discovered approximate Pareto frontiers (line 11), collects the new nodes within that layer,  \revision{denoted as $\mathcal{S}_k$} (line 12), and breaks if the set is not empty (lines 13-14). 
We only collect new nodes from the nearest Pareto frontier rather than from the entire set to avoid the neighbor explosion issue and improve efficiency. If no new nodes exist in $Res$, then we break the loop (lines 15-16). This is because no new nodes within $Res$ can be used to improve the results further. Otherwise, we add the neighbors of the new nodes to $Res$, mark these new nodes as old nodes, and mark their neighbors as visited (lines 17-22). After that, we  \revision{optimize} the candidate set $Res$ by finding the multi-layer Pareto frontiers within $Res$ using $FindPF$,  \revision{and then updating $Res$ accordingly} 
(line 23). 


\noindent\textbf{Complexity Analysis.} 
The dominant factor in the GPS algorithm's time complexity is the search path length, which determines the number of evaluated objects. In an extreme scenario where each layer of the Pareto frontier contains only one node, the GPS algorithm becomes equivalent to the greedy search algorithm with the longest search path length. In other cases, the GPS algorithm performs a greedy search for different $\alpha$ values. Given the bounded size of the candidate set, the search path length is shorter. Therefore, the time complexity of the GPS algorithm is $O(\log(N))$, the same as the greedy search.




\subsection{Dynamic Edge Pruning Strategy}
\label{sec:D-RNG}



As discussed before, we aim to dynamically utilize the RNG’s pruning strategy to prune edges. 
To achieve this, we take $\alpha$ into account the pruning strategy of the RNG, transforming the pruning condition of edge $(x, y)$ into the following two formulas:



\begin{equation}
\begin{aligned}
\alpha \times \delta_e(x, z) + (1 - \alpha) \times \delta_s(x, z) < \\ 
\alpha \times \delta_e(x, y) + (1 - \alpha) \times \delta_s(x, y)
\end{aligned}
\label{eq:rng-hvq-1}
\end{equation}
\begin{equation}
\begin{aligned}
\alpha \times \delta_e(y, z) + (1 - \alpha) \times \delta_s(y, z) < \\
\alpha \times \delta_e(x, y) + (1 - \alpha) \times \delta_s(x, y)
\end{aligned}
\label{eq:rng-hvq-2}
\end{equation}
If both Equation~\ref{eq:rng-hvq-1} and Equation~\ref{eq:rng-hvq-2} hold for any $\alpha \in [0, 1]$, then the edge $(x, y)$ will be pruned due to the presence of node $z$ according to the pruning strategy of RNG, as it becomes the longest edge in the triangle $(x, y, z)$. However, the two equations may be satisfied for some values of $\alpha$ and not for others, which makes the pruning process challenging.

\begin{table}[!t]
\caption{Examples.} 
\centering
\begin{tabular}{p{0.1\textwidth} C{0.1\textwidth} C{0.1\textwidth} C{0.1\textwidth} C{0.1\textwidth} C{0.1\textwidth} C{0.1\textwidth}}
\toprule
\scalebox{0.95}{Examples} & \scalebox{0.9}{$\delta_s(x, y)$} & \scalebox{0.9}{$\delta_e(x, y)$} & \scalebox{0.9}{$\delta_s(x, z)$} & \scalebox{0.9}{$\delta_e(x, z)$} & \scalebox{0.9}{$\delta_s(y, z)$} & \scalebox{0.9}{$\delta_e(y, z)$}  \\ 
\midrule
\scalebox{0.9}{$(x_1, y_1, z_1)$} & 0.3 & 0.4 & 0.8 & 0.9 & 0.1 & 0.7 \\ 
\scalebox{0.9}{$(x_2, y_2, z_2)$} & 0.5 & 0.7 & 0.2 & 0.4 & 0.3 & 0.5 \\ 
\scalebox{0.9}{$(x_3, y_3, z_3)$} & 0.2 & 0.6 & 0.4 & 0.5 & 0.3 & 0.4 \\ 
\bottomrule
\end{tabular}
\label{tab:example}
\end{table}

\noindent{\textbf{Example 1:}} Table~\ref{tab:example} presents several examples to illustrate this challenge. In the first example, $(x_1, y_1, z_1)$, Equation~\ref{eq:rng-hvq-1} does not hold for any $\alpha$ value. This means that the edge $(x, y)$ will not be pruned due to the presence of node $z$ according to RNG's pruning strategy. In the second example, $(x_2, y_2, z_2)$, both equations are satisfied for any $\alpha \in [0, 1]$, indicating that the edge $(x_2, y_2)$ will be consistently pruned by node $z_2$, regardless of the $\alpha$ value. In the third example, $(x_3, y_3, z_3)$, the first equation holds for $\alpha \in [\frac{2}{3}, 1]$ and the second equation holds for $\alpha \in [\frac{1}{3}, 1]$. This means that the edge $(x_3, y_3)$ will be pruned due to the presence of node $z$ when $\alpha \in [\frac{2}{3}, 1]$, as both conditions are satisfied in this range.

Example~1 shows that the RNG's pruning strategy is effective in some $\alpha$ cases, but not in others. We formalize this property into the following lemma.
\vspace*{-0.5em}
\begin{lemma}
\label{lemma:drng}
If Equation~\ref{eq:rng-hvq-1} holds for $\alpha \in r^z_1$ and Equation~\ref{eq:rng-hvq-2} holds for $\alpha \in r^z_2$, where $r^z_1, r^z_2 \subseteq [0, 1]$, then the edge $(x, y)$ will be pruned due to the presence of node $z$ according to RNG's pruning strategy when $\alpha \in r^z_1 \cap r^z_2$. 
\end{lemma}
\vspace*{-0.5em}
Due to page limitations, the proof is provided in the appendix. Next, we demonstrate how to compute $r^z_1$ and $r^z_2$. First, 
Equations~2 and 3
can be transformed into the following two formulas, respectively:



\begin{equation}
\begin{aligned}
\alpha \times (\delta_e(x, z) - \delta_e(x, y) + \delta_s(x, y) - \delta_s(x, z)) < \\
\delta_s(x, y) - \delta_s(x, z)
\end{aligned}
\label{eq:rng-hvq-3}
\end{equation}
\begin{equation}
\begin{aligned}
\alpha \times (\delta_e(y, z) - \delta_e(x, y) + \delta_s(x, y) - \delta_s(y, z)) < \\
\delta_s(x, y) - \delta_s(y, z)
\end{aligned}
\label{eq:rng-hvq-4}
\end{equation}
In Equation~\ref{eq:rng-hvq-3}, the range of $\alpha$ that satisfies the inequality, denoted as $r_1^z$, is determined by the value and sign of the distance differences, which can be categorized into four cases based on their signs:


\noindent\textbf{Case 1:} If $\delta_s(x, y)-\delta_s(x, z) > 0$ and $\delta_e(x, z)-\delta_e(x, y)+\delta_s(x, y)-\delta_s(x, z) > 0$, then the inequality is satisfied for the range $r_1^z = \left[0, \min\left(1, \frac{\delta_s(x, y) - \delta_s(x, z)}{\delta_e(x, z) - \delta_e(x, y) + \delta_s(x, y) - \delta_s(x, z)}\right)\right]$.

\noindent\textbf{Case 2:} If $\delta_s(x, y) - \delta_s(x, z) < 0$ and $\delta_e(x, z) - \delta_e(x, y) + \delta_s(x, y) - \delta_s(x, z) \geq 0$, then the inequality cannot be satisfied for any $\alpha \in [0, 1]$, resulting in $r_1^z = \emptyset$.

\noindent\textbf{Case 3:} If $\delta_s(x, y) - \delta_s(x, z) > 0$ and $\delta_e(x, z) - \delta_e(x, y) + \delta_s(x, y) - \delta_s(x, z) \leq 0$, then the inequality can be satisfied for any $\alpha \in [0, 1]$, resulting in $r_1^z = [0, 1]$. 

\noindent\textbf{Case 4:} If $\delta_s(x,y) - \delta_s(x,z) < 0$ and $\delta_e(x,z) - \delta_e(x,y) + \delta_s(x,y) - \delta_s(x,z) < 0$, then the inequality is satisfied for the range $r_1^z = \left[\min\left(1, \frac{\delta_s(x,y) - \delta_s(x,z)}{\delta_e(x,z) - \delta_e(x,y) + \delta_s(x,y) - \delta_s(x,z)}\right), 1\right]$.

\noindent\textbf{Example 2:} The example $(x_1, y_1, z_1)$ falls under case 2, resulting in $r_1^z = \emptyset$. The example $(x_2, y_2, z_2)$ falls under case 3, leading to $r_1^z = [0, 1]$. The example $(x_3, y_3, z_3)$ falls under case 4, $r_1^z = [\frac{2}{3}, 1]$. 

For Equation~\ref{eq:rng-hvq-4}, such range can computed similarly, denoted as $r_2^z$. 
Thus, the pruning range of edge $(x, y)$ due to the presence of node $z$ is the intersection of $r_1^z$ and $r_2^z$, denoted as $r^z = r_1^z \cap r_2^z$. The active range $u$ of edge $(x, y)$ for node $z$ is then the complement of $r^z$, given by $u = [0, 1] \setminus r^z$.





\begin{algorithm}[!t]
    \caption{\revision{Dynamic Edge Pruning Strategy}}
    \small
    \label{alg:DRNG-prune}
    \KwIn{A candidate set $CS$ consisting of approximate Pareto frontiers, maximum edges $M$, a threshold value $th$ to prune edges with a small use range}  
    \KwOut{Neighbor Set $NS$}
    \BlankLine    
    \SetKwFunction{DRNGPrune}{\textsc{DRNGPrune}}
    \SetKwProg{Fn}{Procedure}{:}{}
    \Fn{\DRNGPrune{$CS$, $M$, $th$}}{
        Initialize the Neighbor Set $NS = \emptyset$\;        
        \ForEach{$PF_i \in CS$}{
            \ForEach{$x \in PF_i$}{
                $r^x \leftarrow \emptyset$\;
                \ForEach{$y \in NS$}{
                    Compute $r^y$\;
                    $r[x] \leftarrow r[x] \cup r^y$\;
                }
                $u \leftarrow [0, 1] \setminus r[x]$ \;
                \If{$|u| \geq th$}{
                    $NS.\text{add}({x, u})$\;
                }                
            }
            \If{$|NS|\geq M$}{
                \textbf{break}\;  
            }
        }
        \Return $NS$\;
    }
\end{algorithm}

\noindent\textbf{\method's Pruning Strategy.} Using the approximate Pareto frontiers derived by the GPS algorithm as the candidate set $CS$, we can apply this dynamic edge pruning strategy to it. As shown in Algorithm~\ref{alg:DRNG-prune}, we first initialize the neighbor set $NS$ as an empty set (line 2). Then we sequentially traverse each layer of the Pareto frontier from nearest to farthest, gradually adding nodes to the $NS$ (lines 3--13). Specifically, for each new node $x$, we compute its pruning range $r^y$ with respect to each previously added node $y$ and take their union $r[x]$ as the final pruning range (lines 6--8), ensuring that the new edge will not be pruned by any previously added edges. The active range $u$ for this edge is the complement of $r[x]$ (line 9). To avoid maintaining edges that are not useful in most cases (e.g., $u = [0, 0.05]$), we set a threshold value $th$ to further prune such edges (lines 10--11). Once we obtain $M$ edges, we break the loop and return $NS$ as the final neighbor set (lines 12--13).

\revision{
However, our pruning strategy does not apply to multi-vector queries with more than two vectors, as the active range becomes a hyperplane in 2D space in these cases, which is challenging to compute, store, and utilize. We leave this 
for future work.}
\vspace*{-0.5em}
\begin{lemma}
\label{lemma:drng-nearest-neighbor}
By considering all objects in the dataset as candidate neighbors for each node and applying the dynamic edge pruning strategy to obtain dynamic edges, the constructed graph becomes a Relative Neighborhood Graph~\cite{fu2019fast} for any $\alpha$ value. This ensures that the nearest neighbor can always be found for the query using the greedy search algorithm. 
\end{lemma}
\vspace*{-0.5em}
Due to page limitations, the proof is provided in the appendix.
\subsection{Index Construction}
\label{sec:construction}

\noindent\textbf{Edge Seed Acquisition.} 
Another challenge discussed in Section~\ref{sec:intro} is how to choose the start nodes (also known as the seed) for the graph index. 
Existing graph-based ANNS methods, such as HNSW~\cite{malkovEfficientRobustApproximate2020}, randomly select a subset of nodes as the seed set, which forms the upper layers. To reduce search path length and improve search efficiency, one approach~\cite{fu2019fast} sets the node closest to the graph's center as the seed. 
However, for the \hvq problem, as $\alpha$ changes, using a single centroid or a small number of random nodes as the seed results in reduced performance (as to be shown in Section~\ref{sec:exp-ablation-study}). A straightforward method to maintain performance is to sample more random nodes or choose multiple centroids as seeds for different $\alpha$ values, but this reduces the efficiency during the query phase due to the multiple start nodes.

Figure~\ref{fig:seed} illustrates this issue, where the red nodes represent the potential multiple centers. At query time, the greedy search algorithm used by graph-based ANNS indexes maintains two sets: a min-heap candidate set $\mathcal{S}$ and a max-heap result set $\mathcal{R}$, as detailed in Section~\ref{sec:graph-anns}. The algorithm iteratively fetches the object with the minimum distance in $\mathcal{S}$ and adds its neighbors into $\mathcal{S}$ and $\mathcal{R}$. When we use multiple centroids as seeds, they all have similar distances to the query, as shown in Figure~\ref{fig:seed} by $\mathcal{S}_2$. This can lead to multiple parallel searches, causing many intermediate nodes to be visited repeatedly, thus reducing efficiency.
To address this, we propose a novel edge seed acquisition method. The core idea of this method is to choose the nodes farthest from the center under varying $\alpha$ value as the seeds, which are located at the edge of the graph. Figure~\ref{fig:seed} illustrates this method, where the green nodes indicate the edge seed. These edge seeds are much farther from the query. Therefore, during the greedy search process, these distant seeds will remain at the bottom of the candidate set $\mathcal{S}_1$, and their neighbors will not be considered and evaluated. From a high-level perspective, the edge seed method allows us to adaptively start the search from the edge node closest to the query while ignoring other distant edge nodes.



To obtain the edge seed, we employ an efficient yet highly effective method. Specifically, we first calculate the centroid $c$ of $D$, where $c.e = \text{avg}(x.e)$ and $c.s = \text{avg}(x.s)$ for all $x \in D$. Then we maintain the inverse Pareto frontier of the centroid $c$, which consists of the nodes that do not dominate any other nodes based on the distance from the centroid, i.e., the most distant nodes from the centroid under varying $\alpha$. The algorithm can be easily adapted from existing algorithm for finding two-dimensional skyline~\cite{borzsony2001skyline}
, so the details are not provided here. The time complexity of updating the seed set $ep$ is $O(|ep|\log(|ep|) + |ep|)$ since we only need to maintain one layer of inverse Pareto frontier. The inverse Pareto frontier often contains only a dozen nodes, so the time complexity for updating is negligible.

\begin{algorithm}[t]
    \caption{Index Construction}
    \small
    \label{alg:index-construction}
    \KwIn{A dataset $D$, candidate pool size $ef_{construction}$, maximum edges per node $M$, a threshold value $th$ to prune edges with a small use range}    
    \KwOut{$G(V, E, ep)$, where $ep$ denotes the seed set}
    \BlankLine    
    \SetKwFunction{DEGBuild}{\textsc{DEGBuild}}
    \SetKwProg{Fn}{Procedure}{:}{}
    \Fn{\DEGBuild{$D$, $ef_{construction}$, $M$, $th$}}{
        calculate the centroid $c$ of $D$\;        
        Initialize graph $G(V, E, ep)$, $V = \{x_0\}$, $E = \emptyset$, $ep = \{x_0\}$\;        
        \ForEach{$x \in D \setminus \{x_0\}$}{
            $V \leftarrow V \cup \{x\}$\;
            $CS \leftarrow \GPS(G, x, ep, ef_{construction})$\;
            $NS(x) \leftarrow \DRNGPrune(CS, M, th)$\;
            $E(x) \leftarrow NS(x)$\;                                 
            \ForEach{$y \in NS(x)$ }{
                $E(y) \leftarrow \DRNGPrune(E(y)\cup \{x\}, M, th)$\; 
            }
            update $ep$ as the inverse Pareto frontier of $c$\;
        }
        \Return $G$\;
    }
\end{algorithm}

\noindent\textbf{Summary.} Based on the modules proposed above, we summarize the index construction process. The details of the index construction phase are provided in Algorithm~\ref{alg:index-construction}. Specifically, we first calculate the centroid $c$ of $D$ (line 2). 
Then we initialize the start node as the first node $x_0$ (line 3) and construct the graph by iteratively inserting new nodes (lines 4-11). For each newly inserted node $x$, we employ the GPS algorithm to search for approximate Pareto frontiers (line 6) and obtain the final neighbor set $NS(x)$ of node $x$ among the approximate Pareto frontiers using Algorithm~\ref{alg:DRNG-prune} (line 7). The derived neighbor set $NS(x)$ is treated as $x$'s edges in $G$ (line 8). Similar to previous studies~\cite{malkovEfficientRobustApproximate2020}, we also attempt to add reverse edges by trying to include $x$ in $E(y)$, where $y \in NS(x)$ (lines 9-10). Finally, we update the edge seed set by determining whether $x$ can be added to the inverse Pareto frontier of the centroid (line 11).

\subsection{Search Algorithm}
\label{sec:search}
We next present the search algorithm of \method. It is based on the greedy search algorithm~\cite{wang2021comprehensive}, with two modifications. Firstly, it dynamically skips edges whose active ranges do not intersect with the query's $\alpha$ value. Secondly, we introduce an early stopping mechanism. When calculating the hybrid distance, we first compute one of the individual vector distances and compare it to the threshold distance in the greedy search algorithm to determine whether to skip this node early.

Specifically, the algorithm first initializes an empty min-heap $\mathcal{S}$ as the candidate set, an empty max-heap $\mathcal{R}$ as the result set. Next, the seed are added to $\mathcal{S}$, $\mathcal{R}$. In each iteration, the object with the smallest distance in $\mathcal{S}$ is fetched. If its distance to the query is larger than the maximum distance in $\mathcal{R}$, the loop breaks, and the top $k$ objects in $\mathcal{R}$ are returned. Otherwise, its neighbors are checked, with some being skipped based on the active ranges and the query $\alpha$ value. If the distance of a neighbor to the query is smaller than the maximum distance in $\mathcal{R}$, the neighbor is pushed into $\mathcal{S}$ and $\mathcal{R}$. Here, the early termination mechanism is used for acceleration. If the size of $\mathcal{R}$ exceeds $ef_{search}$, the object with the maximum distance is popped from $\mathcal{R}$.

\noindent\textbf{Complexity Analysis.} Here, we analyze the time and space complexity of \method. For space complexity, the main difference between \method and previous graph-based ANNS indexes is the active range $u$ stored for each edge. Compared to the high-dimensional vectors stored in memory, this additional storage is negligible. As for time complexity, 
during the query phase, 
\method dynamically skips some edges, while the remaining edges satisfy the RNG property. Therefore, the search time complexity remains the same as that of previous RNG-based ANNS indexes~\cite{wang2021comprehensive}, which is $O(\log(N))$.

\section{Experiments}
\label{sec:exp}


\subsection{Evaluation Setup}
\label{sec:exp-setup}

\begin{table}[!t]
    \centering
    \small
    \caption{Datasets Statistics.}
    \vspace*{-1em}
    \label{tab:dataset}
    \renewcommand{\arraystretch}{1.2} 
    \setlength{\arrayrulewidth}{1pt}
    \begin{tabularx}{0.9\textwidth}
    {|>{\centering\arraybackslash}p{2cm}|>{\centering\arraybackslash}p{1.5cm}|>{\centering\arraybackslash}p{0.8cm}|>{\centering\arraybackslash}p{0.8cm}|>{\centering\arraybackslash}p{0.8cm}|>{\centering\arraybackslash}X|}
    \hline
    Dataset & $|D|$ & $d$ & $m$ & $|Q|$ & Type  \\
    \hline
    \textsf{OpenImage} & 507,444 & 768 & 768 & 1,000 & Text, Image \\
    \hline
    \textsf{Ins-SG}  & 1,000,000 &  768 & 2 & 1,000 & Text, Coordinate \\ 
    \hline
    \textsf{Howto100M} & 1,238,875& 1,024 & 768 & 1,000 & Text, Video \\    
    \hline
    \textsf{CC3M} & 3,131,153 &  768 & 768 & 1,000 & Text, Image \\
    \hline
    \textsf{Twitter-US} & 10,000,000 &  768 & 2 & 1,000 & Text, Coordinate \\
    \hline
    \end{tabularx}
    \vspace*{-1em}
\end{table}

\noindent\textbf{Datasets.} Our experiments are conducted on five real-world datasets: \textsf{OpenImage}, \textsf{Ins-SG}, 
\textsf{Howto100M}, \textsf{CC3M}, and \textsf{Twitter-US}. The statistics of datasets are listed in Table~\ref{tab:dataset}. Details of each dataset are stated as follows.
\begin{itemize}[leftmargin=*, topsep=0pt]
\item \textsf{OpenImage}~\cite{OpenImagesLocNarr}: The \textsf{OpenImage} dataset\footnote{\url{https://google.github.io/localized-narratives/}} is an open benchmark for object detection, image classification, and visual relationship detection. 
It comprises 500K training images and 41K validation images. Each image is paired with localized narratives provided by annotators. Images are converted into 768-dimensional vectors ($o.e$) using ViT~\cite{dosovitskiy2020image}, and localized narratives are transformed into 768-dimensional vectors ($o.s$) using BERT~\cite{devlin2018bert}. The training set is used as the 
database. The query set $Q$ ($|Q|$ = 1,000 ) is 
randomly selected from the validation set.

\item \textsf{Ins-SG}: The \textsf{Ins-SG} dataset is a real dataset that contains 1 million Instagram posts from Singapore. Each post has a  geo-location and image. The images are transformed into 
vectors using ViT. The query set $Q$ consists of another 1,000 collected posts. 

\item \textsf{Howto100M}~\cite{miech19howto100m}: The \textsf{Howto100M} dataset\footnote{\url{https://www.di.ens.fr/willow/research/howto100m/}} is an open benchmark for learning text-video embeddings.
It includes 136 million video clips,
covering 23,000 activities in areas such as cooking, handcrafting, and fitness. Each video is paired with subtitles automatically downloaded from YouTube. We downloaded S3D~\cite{miech2020end} video embeddings from its official website and transformed textual descriptions into embeddings using BERT. We randomly sampled 1,000 video-text pairs 
to form the query set $Q$.


\item \textsf{CC3M}~\cite{sharma2018conceptual}: The Conceptual Captions dataset\footnote{\url{https://github.com/google-research-datasets/conceptual-captions}} is an open benchmark for training and evaluating visual-language models. It comprises 3.3M image-text pairs. 
randomly selected 1,000 image-text pairs as the query set, and treated the remaining image-text pairs as the database. Each Image and its attached text are 
transformed into vectors using ViT and BERT, respectively.

\item \textsf{Twitter-US}: The \textsf{Twitter-US} dataset is generated from 10 million real tweets in the USA. All of them contain geo-locations and text descriptions. The text descriptions are transformed into vectors using BERT. The query set $Q$ consists of 1,000 real tweets collected together with the data. This dataset is an order of magnitude larger than those used by other memory-based graph ANNS index studies~\cite{wang2021comprehensive}. Therefore, we conduct the scalability study on this dataset.

\end{itemize}

\noindent\textbf{Evaluation metrics.} Exiting ANNS studies~\cite{liApproximateNearestNeighbor2020,wang2021comprehensive} typically use recall rate $Recall@k=\frac{\mathcal{R}\cap \tilde{\mathcal{R}}}{k}$ to evaluate the accuracy of search results and queries per second QPS $= \frac{\#q}{t}$ to evaluate the search's efficiency. Here, $\mathcal{R}$ represents the result set retrieved by the index, $\tilde{\mathcal{R}}$ denotes the 
result set returned by a brute-force search, and $|\mathcal{R}| = |\tilde{\mathcal{R}}| = k$. QPS is the ratio of number of queries ($|Q|$) to search time ($t$); i.e., QPS$ = \frac{|Q|}{t}$~\cite{fu2019fast}. In this work, we use recall@10 and QPS as the evaluation metrics.


\noindent\revision{
\textbf{Baselines.} In addition to the two existing baselines discussed in Section~\ref{sec:motivations}, we 
consider a baseline called Overlay, and an ideal method called Oracle~\footnote{We would like to thank the anonymous reviewers for suggesting the two methods}.
}
\begin{itemize}[leftmargin=*, topsep=0pt]
    \item \revision{\textsf{Overlay (abbr. O)}: This method constructs five different graph-based ANNS indexes by setting $\alpha$ in the hybrid distance to $0.1$, $0.3$, $0.5$, $0.7$, $0.9$, respectively. These separate graph-based indexes are then overlaid into a single graph-based index by merging their edges accordingly, with each edge assigned a value representing the $\alpha$ under which it was constructed. During the query phase, 
    rather than traversing all edges, it only routes the edges with values closest to the query’s $\alpha$ and ignores others. This actually restricts the search to the corresponding sub-index, i.e., we use the sub-index built with $q.\alpha=0.1$ to handle queries with $\alpha \in [0, 0.2]$, the sub-index with $\alpha=0.3$ for queries with $q.\alpha \in [0.2, 0.4]$, etc. The overlay operation eliminates the need to store five separate indexes, thereby reducing memory usage. The accuracy-efficiency trade-off is controlled by the search algorithm's hyperparameter (e.g., $ef_{search}$). We integrate this method with the HNSW index, denoted as \textsf{HNSW}$_{\textsf{O}}$. }  
    \item \revision{\textsf{Oracle (abbr. Or)}: This method represents the ideal scenario where a separate graph-based ANNS is built for every possible $\alpha$ value, with searches conducted on the corresponding index. Although this method is not feasible in practical scenarios since $q.\alpha$ is unknown beforehand and can be an arbitrary value within $[0, 1]$, we use it to illustrate the performance gap between our proposed method and the ideal case. Specifically, we implement this approach using the HNSW index, denoted as \textsf{HNSW}$_{\textsf{Or}}$. To make the comparison feasible, we select five $q.\alpha$ values (0.1, 0.3, 0.5, 0.7, and 0.9), constructing a separate HNSW index for each, with $M = 40$ and $ef_{construction} = 200$. These five $\alpha$ values are used to generate five test query sets.
    We compare \method with \textsf{HNSW}$_{\textsf{Or}}$ and other baselines on these query sets.}
    
\end{itemize}


\begin{figure*}[!t]
\centering
\vspace*{-1.5em}
\subcaptionbox{\revision{OpenImage}\label{fig:exp-openimage}}{
\includegraphics[width=1.0\textwidth]{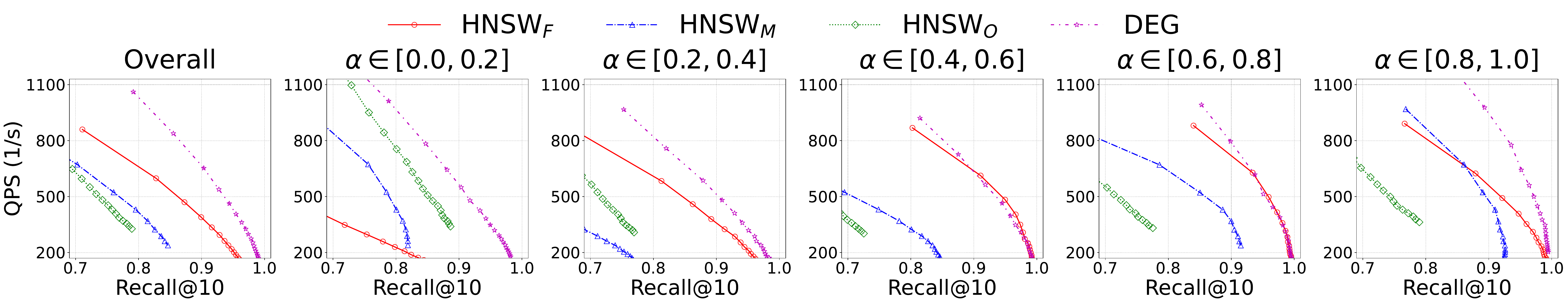}
\vspace*{-0.5em}
}
\subcaptionbox{\revision{Ins-SG}\label{fig:exp-sg-ins}}{
\includegraphics[width=1.0\textwidth]{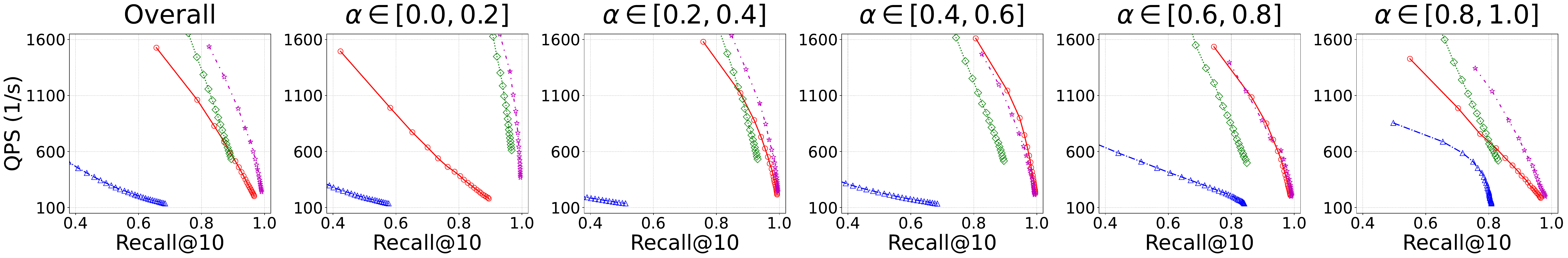}
\vspace*{-0.5em}
}
\subcaptionbox{\revision{Howto100M}\label{fig:exp-howto100m}}{
\includegraphics[width=1.0\textwidth]{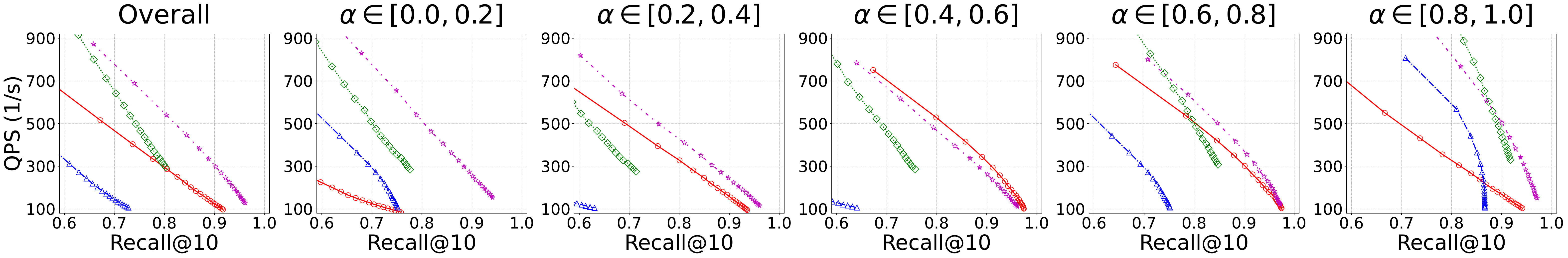}
\vspace*{-0.5em}
}
\subcaptionbox{\revision{CC3M}\label{fig:exp-cc3m}}{
\includegraphics[width=1.0\textwidth]{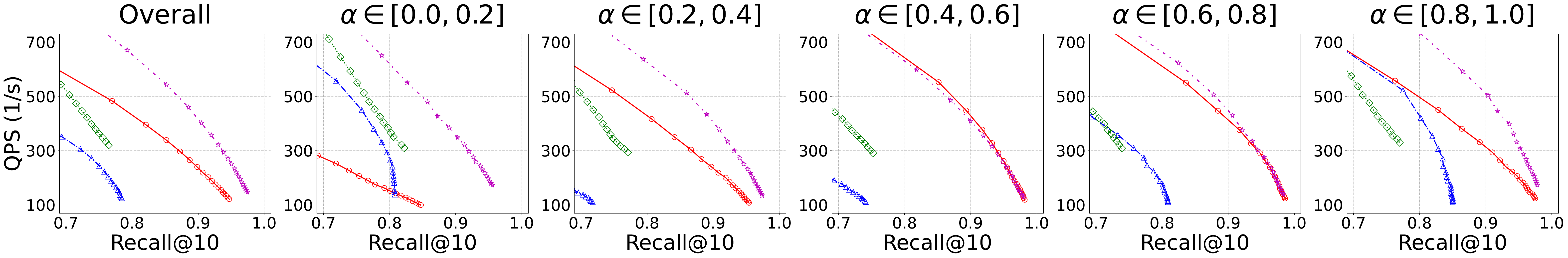}
\vspace*{-0.5em}
}
\vspace*{-1em}
\caption{\revision{The accuracy-efficiency trade-off results (upper and right is better).}}
\vspace*{-1.5em}
\label{fig:exp-acc-eff}
\end{figure*}

\begin{figure*}[!t]
\centering
\subcaptionbox{\revision{OpenImage}\label{fig:exp-openimage-oracle}}{
\includegraphics[width=1.0\textwidth]{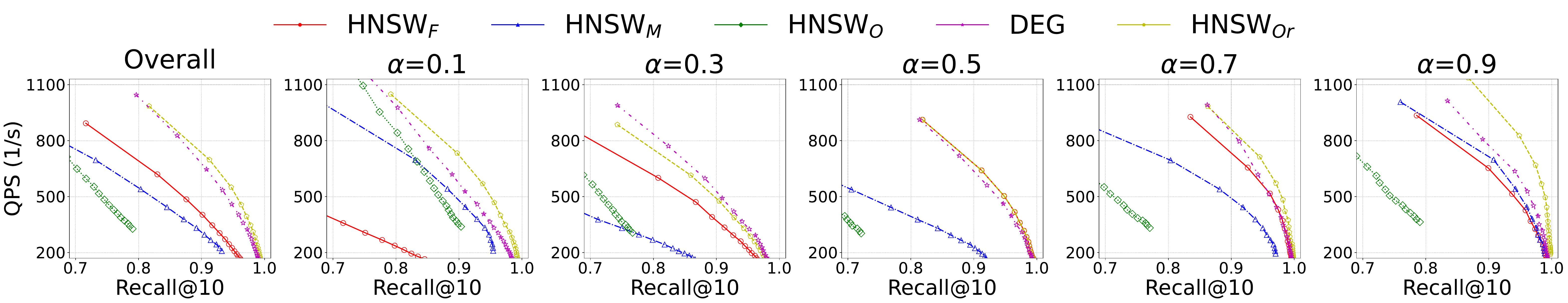}
\vspace*{-0.5em}
}
\vspace*{-1em}
\caption{\revision{The accuracy-efficiency trade-off results compared to \textsf{HNSW}$_{\textsf{Or}}$ (upper and right is better).}}
\vspace*{-1.5em}
\label{fig:exp-acc-eff-oracle}
\end{figure*}

\noindent\textbf{Parameter Settings.} 
The three key parameters of HNSW, namely the candidate set size $ef_{construction}$, the maximum number of edges per node $M$, and the search set size $ef_{search}$, are set to 200, 40, and 10 by default for baselines \textsf{HNSW}$_{\textsf{F}}$ and \textsf{HNSW}$_{\textsf{M}}$, with other parameters set as recommended in the previous study~\cite{malkovEfficientRobustApproximate2020}. \revision{For baseline \textsf{HNSW}$_{\textsf{O}}$, if the maximum number of edges per sub-index node is set to be $M$, the construction time and memory costs can be up to five times higher than other methods, 
leading to unfair comparison. 
Therefore, we set the maximum number of edges per node in each sub-index to 10 by default, so that each node in the overlaid HNSW index has a maximum of 50 edges, and thus the total number of edges in the overlaid graph index is comparable to others. Accordingly, we set the default value of $ef_{construction}$ for each sub-index to 50, which is five times the value of $M$, as applied previously. All other parameters for \textsf{HNSW}$_{\textsf{O}}$ are set the same as other baselines. Experiments on varying $M$ and $ef_{construction}$ will be presented later in Section~\ref{sec:exp-para-sen}. } 
For \textsf{HNSW}$_{\textsf{F}}$ \revision{and \textsf{HNSW}$_{\textsf{O}}$}, we vary the hyperparameter $ef_{search}$ from 10 to 200 in steps of 10 to control the accuracy-efficiency trade-off. For \textsf{HNSW}$_{\textsf{M}}$, we vary the hyperparameter $k'$, the number of objects to be retrieved from each index, and then merged and reranked, from 10 to 200 in steps of 10 to control the accuracy-efficiency trade-off, with $ef_{search}$ always set equal to $k'$. 
{By default, we set $M$ to 40 and $ef_{construction}$ to 200 for \method, aligning with \textsf{HNSW}$_{\textsf{F}}$. When \textsf{HNSW}$_{\textsf{F}}$'s hyperparameters are adjusted, \method is modified to the same value accordingly to ensure consistency and fair comparison unless otherwise specified.} The threshold value $th$ is consistently set to $0.1$. The parameter $\alpha$ in Equation~\ref{eq:hybird_distance} allows to set preferences between different modalities. \revision{
To evaluate the capability of methods to adapt to different query $\alpha$ values, we divide the $\alpha$ range into five intervals: [0, 0.2], [0.2, 0.4], [0.4, 0.6], [0.6, 0.8], and [0.8, 1]. We generate five random $\alpha$ values for each test query, one from each interval, producing five different query sets. Each set includes all test queries with $\alpha$ values from one specified interval. 
We report the overall average results of all sets for a comprehensive comparison. We also report the average results for each set for a detailed comparison. 
}

\noindent\textbf{Implementations.} 
The baselines and  \method\footnote{The code is available at \url{https://github.com/Heisenberg-Yin/DEG}.} are all implemented in C++. The implementations of HNSW\footnote{\url{https://github.com/Lsyhprum/WEAVESS/}} and R-tree\footnote{\url{https://github.com/nushoin/RTree}} are sourced from publicly available code repositories. \revision{
\method and the baselines are implemented by following previous experimental evaluations\footnotemark[5], excluding SIMD, pre-fetching, and other hardware-specific optimizations. Although these optimizations can speed up index construction significantly, they are hardware-specific and could introduce unfairness in comparison. }
Our default experimental environment consists of an AMD Ryzen Threadripper PRO 5965WX CPU @ 7.00 GHz and 128GB of memory. 

\subsection{Experimental Results}
\label{sec:exp-results}
\subsubsection{\textbf{Search Performance}}
\label{sec:exp-search-performance}
The accuracy-efficiency trade-off results over the four datasets are shown in Figure~\ref{fig:exp-acc-eff}. We have the following observations.

\noindent
\revision{\textbf{(1) \method demonstrates the best performance compared to the baselines 
on all the datasets.} 
Specifically, compared to the baselines, \method consistently achieves the best overall performance across the four datasets. Additionally, across different $\alpha$ settings, \method consistently delivers the best performance. For example, in the \textsf{OpenImage} dataset, among the baselines, \textsf{HNSW}$_{\textsf{F}}$ performs better when $\alpha \in [0.2, 1]$ and \textsf{HNSW}$_{\textsf{O}}$ excels when $\alpha \in [0, 0.2]$. Our proposed \method matches the performance of \textsf{HNSW}$_{\textsf{F}}$ for $\alpha \in [0.4, 0.6]$ and outperforms the best baseline in all other query $\alpha$ settings. } It is worth noting that \method aims to maintain high performance across different $\alpha$ values rather than outperforming existing state-of-the-art graph-based ANNS indexes. It is as expected that \method has similar performance as \textsf{HNSW}$_{\textsf{F}}$ when $q.\alpha$ \revision{is close to} $0.5$ because \textsf{HNSW}$_{\textsf{F}}$ is constructed with $\alpha=0.5$. 


\noindent\revision{\textbf{(2) The baselines perform well in certain $q.\alpha$ settings but poorly in others, while \method consistently achieves high performance across all $q.\alpha$ settings without significant degradation.} 
The results show that when $q.\alpha$ is close to $0.1$, \textsf{HNSW}$_{\textsf{F}}$ faces significant performance degradation across the four datasets. Similarly, \textsf{HNSW}$_{\textsf{M}}$ shows comparable degradation across the four datasets when $\alpha$ is close to $0.5$. These results validate our analysis in Section~\ref{sec:motivations}. A similar pattern is observed with \textsf{HNSW}$_{\textsf{O}}$, which performs better at $\alpha \in [0, 0.2]$ on the \textsf{OpenImage}, \textsf{Ins-SG}, and \textsf{CC3M} datasets, and excels at $\alpha \in [0.8, 1.0]$ on the \textsf{Howto100M} dataset. 
This pattern may occur because when $q.\alpha$ is close to 0 or 1, the search becomes dominated by a single feature vector. If this modality's feature vector is easier for the graph-based ANNS to capture, the performance will improve. This reasoning is also supported by the performance of \textsf{HNSW}$_{\textsf{M}}$. In settings where \textsf{HNSW}$_{\textsf{O}}$ performs better, \textsf{HNSW}$_{\textsf{M}}$ also shows relatively better performance. For example, on the \textsf{OpenImage} and \textsf{CCM} datasets, for $\alpha$ in [0, 0.2], \textsf{HNSW}$_{\textsf{M}}$ outperforms \textsf{HNSW}$_{\textsf{F}}$, but 
performs worse in other $\alpha$ settings. The \textsf{Ins-SG} dataset is an exception and the reason for this will be explained later. 
Moreover, we also evaluate the performance of \method 
when $\alpha = 0$ or $1$, which reduces to normal vector queries. Experimental results show that \method still delivers 
the best performance compared to the 
baselines for \hvq. Detailed results are provided in the appendix due to page limitations.
} 


\noindent{\textbf{(3) \method maintains similar advantages on larger datasets.} Specifically, the \textsf{CC3M} dataset employs the same embedding techniques as the \textsf{OpenImage} dataset, but it is much larger. \method shows similar advantages over the baselines on both datasets, validating that  \method also performs well on larger datasets.}

\noindent\textbf{(4) \textsf{HNSW}$_{\textsf{M}}$ performs worse on the \textsf{Ins-SG} datasets.} As shown in Figure~\ref{fig:exp-sg-ins}, \textsf{HNSW}$_{\textsf{M}}$ shows significantly worse performance than \textsf{HNSW}$_{\textsf{F}}$ when $\alpha = 0.1, 0.3, 0.5, 0.7$ on the \textsf{Ins-SG} dataset. This is because the \textsf{HNSW}$_{\textsf{M}}$ struggles to retrieve high-quality candidates for reranking. Due to the inherently dense distribution of geographic coordinates compared to high-dimensional vectors, 
hundreds of objects can coexist within a small spatial scale,
making geographically close objects difficult to distinguish from each other, and requiring embedding similarity to further determine the ranking results. \revision{However, both the R-Tree and HNSW used in \textsf{HNSW}$_{\textsf{M}}$ fails to retrieve objects that exhibit similarity across both modalities, thereby resulting in notable performance degradation. 
It is worth mentioning that the throughput of R-Tree and HNSW is comparable, with 1,777 and 1,906 queries per second, respectively, when $k' = 10$. Therefore, the relatively low performance of \textsf{HNSW}$_{\textsf{F}}$ is not due to HNSW having a slower querying speed. 
}

\subsubsection{\revision{\textbf{Performance Gap Analysis with \textsf{HNSW}$_{\textsf{Or}}$.}}} \revision{Here, we compare \method with the \textsf{HNSW}$_{\textsf{Or}}$ and other baselines 
on the \textsf{OpenImage} dataset for $\alpha = 0.1$, $0.3$, $0.5$, $0.7$, and $0.9$. 
The experimental results are shown in Figure~\ref{fig:exp-acc-eff-oracle}. 
\textbf{The results demonstrate that the overall performance of \method is comparable to that of \textsf{HNSW}$_{\textsf{Or}}$ and significantly outperforms other baselines.} 
This suggests that the performance of \method 
is on par with \textsf{HNSW}$_{\textsf{Or}}$.
}



\begin{figure}[!htbp]
\vspace*{-1em}
\centering
\includegraphics[width=0.9\textwidth]{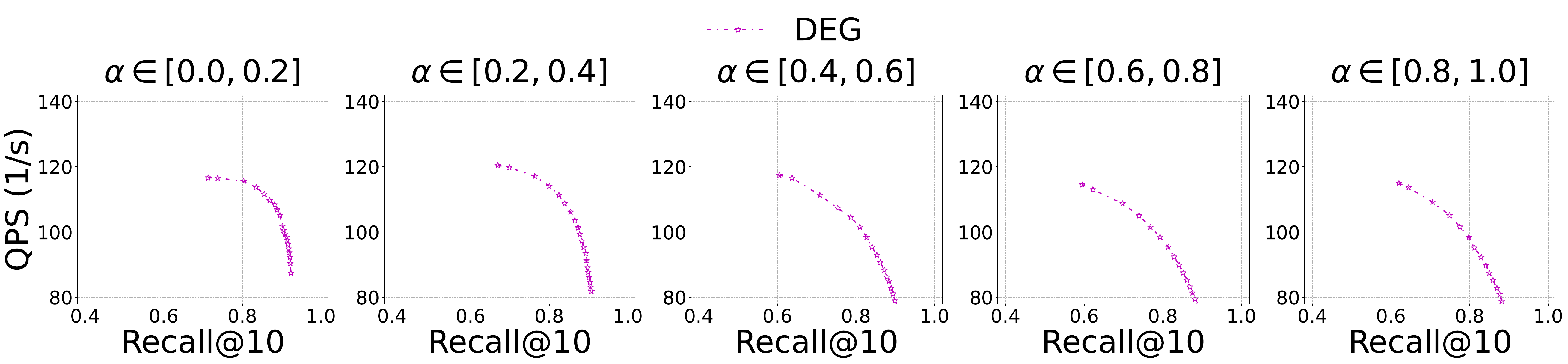}
\vspace*{-1em}
\caption{\revision{Scalability study.}}
\vspace*{-1em}
\label{fig:scalibility-study}
\end{figure}

\subsubsection{\textbf{Scalability Study}}
\label{sec:exp-scalability}
Here, we investigate the index construction cost and search performance of our proposed \method and the baselines on the \textsf{Twitter-us} dataset. The index construction time for \method is 44,492 seconds, approximately 12 hours. However, neither of the baselines completed the index construction within two days. This is consistent with the experimental results of \cite{fu2019fast}, where HNSW could not be constructed on larger datasets and raised an Out-Of-Memory error. We infer that this is due to HNSW's multi-layer mechanism, which causes its construction cost to increase exponentially with the size of the dataset. Therefore, we only report the results for \method, which are shown in Figure~\ref{fig:scalibility-study}. The results show that 
\method exhibits stable performance for varying $\alpha$, consistent with our observations in previous experiments.

\subsubsection{\textbf{Index Cost}}
\label{sec:exp-index-cost}

\begin{figure}[!htbp]
\begin{center}
\subcaptionbox{Construction Time.\label{fig:constrution_time}}{
\includegraphics[width=0.45\columnwidth]{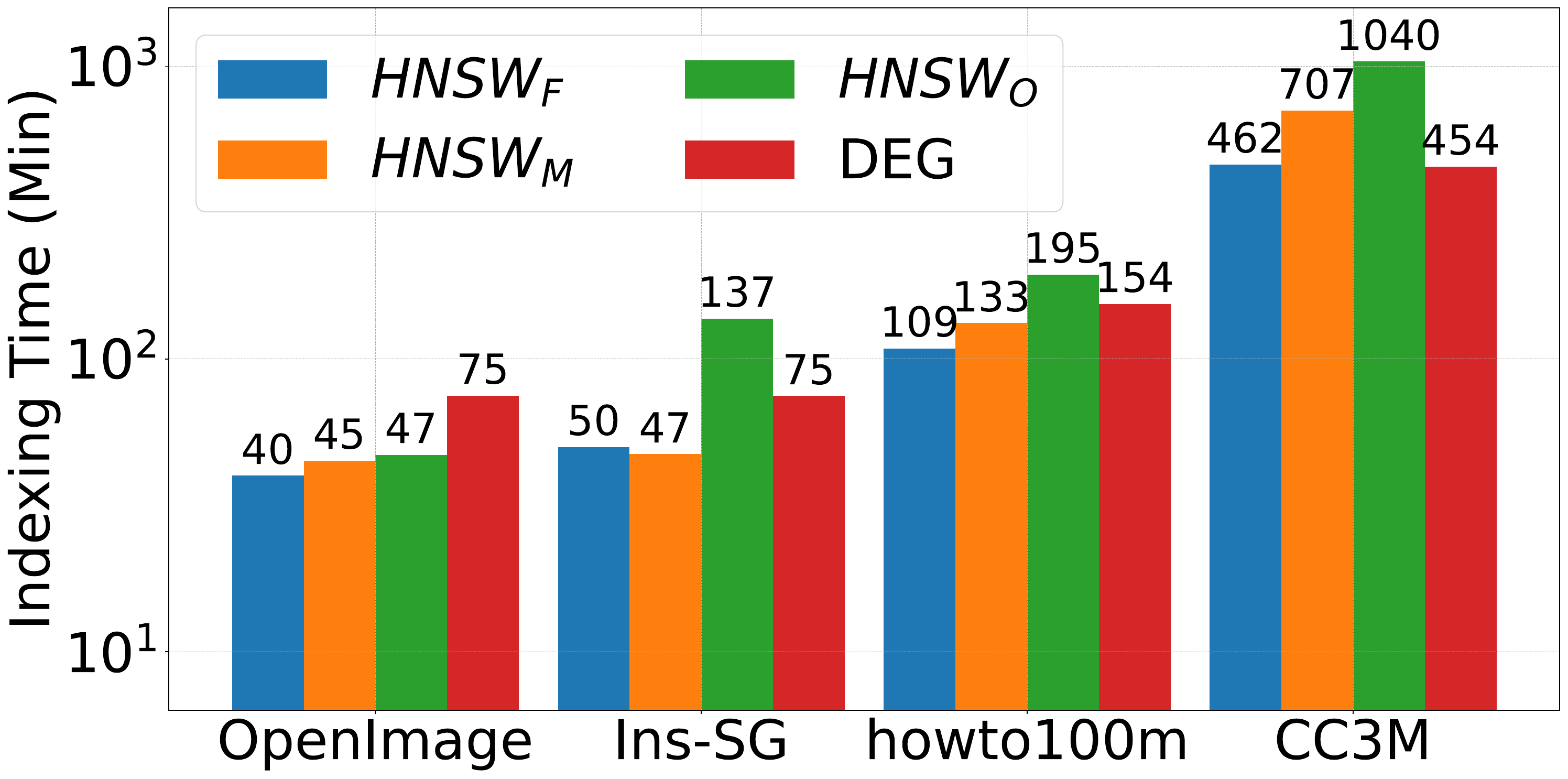}
}
\subcaptionbox{Index Size.\label{fig:memory}}{
\includegraphics[width=0.45\columnwidth]{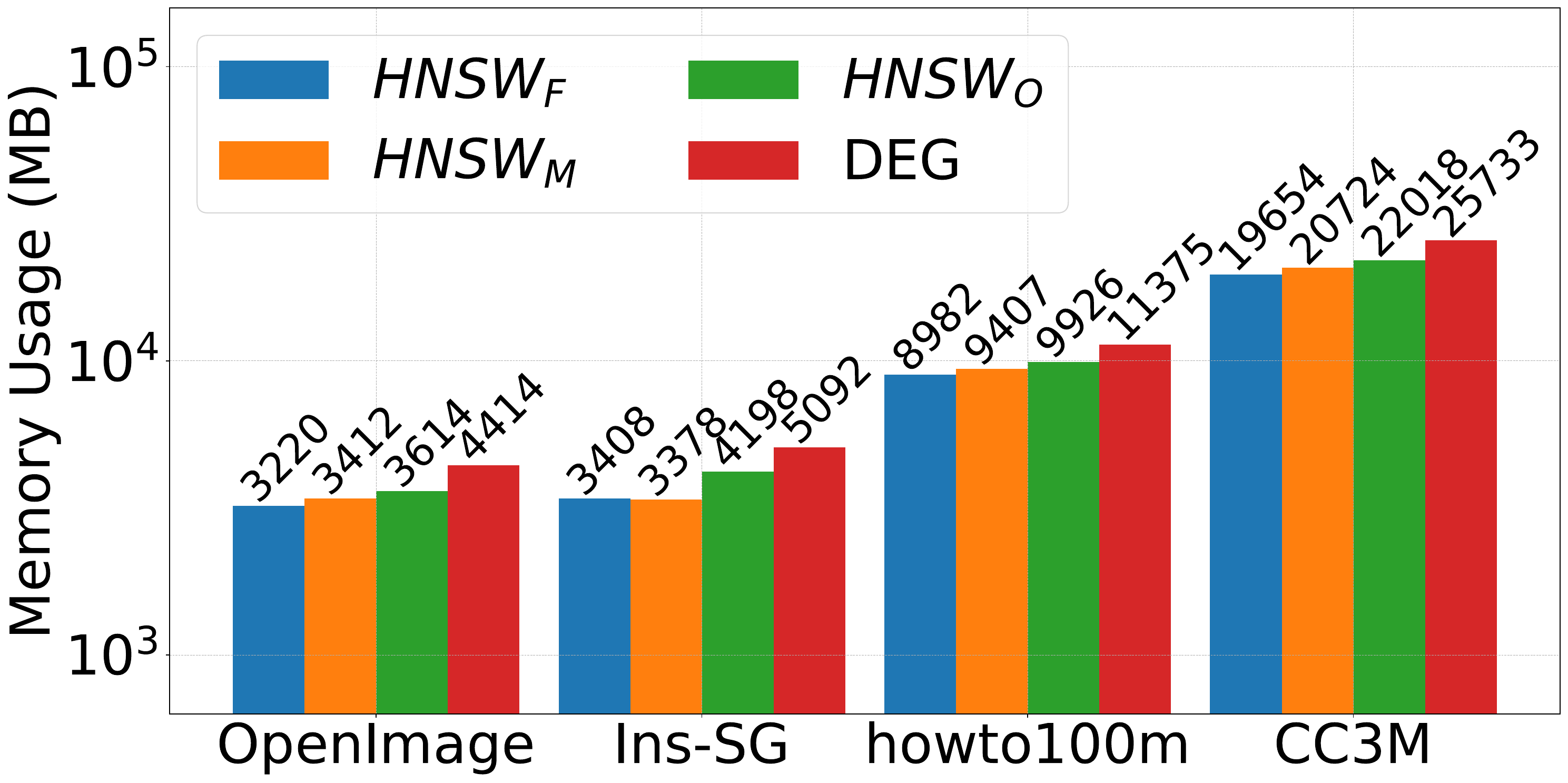}
}
\caption{\revision{The index size and construction time.}}
\label{fig:memory-construction}
\end{center}
\end{figure}

Figure~\ref{fig:constrution_time} illustrates the construction time of \method and the baselines on the four datasets with default parameter settings. \revision{\textbf{
The results show that \method's indexing time is comparable to those of \textsf{HNSW}$_{\textsf{M}}$ and \textsf{HNSW}$_{\textsf{F}}$, and is significantly faster than that of \textsf{HNSW}$_{\textsf{O}}$.}} 
For instance, on the largest dataset \textsf{CC3M}, \method has a comparable time cost to \textsf{HNSW}$_{\textsf{F}}$, \revision{while being 1.6 times faster than \textsf{HNSW}$_{\textsf{M}}$ and 2.3 times faster than \textsf{HNSW}$_{\textsf{O}}$. \textsf{HNSW}$_{\textsf{M}}$ and \textsf{HNSW}$_{\textsf{O}}$ are slower because they  build multiple indexes.} This validates our 
analysis in Section~\ref{sec:search} that the \method's construction time is comparable to previous graph-based ANNS indexes. 
On the \textsf{Ins-SG} dataset, HNSW$_M$ has a slightly shorter construction time than HNSW$_F$. {The reason is two-fold: (1) HNSW$_M$ builds indexes separately, 
computing distances for individual vectors rather than hybrid vectors, resulting in similar construction times across datasets; (2) The construction of R-Tree is faster, leading to slightly quicker times on the \textsf{Ins-SG} dataset but slightly slower on others.} 
\revision{Note that the 
high construction time is a result of our setting, as discussed in Section~\ref{sec:exp-setup}, where hardware-specific optimizations such as SIMD and pre-fetching instructions have been removed. With these optimizations, the construction time for million-scale datasets can be reduced to several minutes.}  Figure~\ref{fig:memory} shows the memory usage of \method and the baselines. \textbf{The results show that \method's memory usage is slightly higher than baseline methods due to its more complex index structure.} However, the high-dimensional vectors still dominate the memory consumption, making the difference negligible 
in its real-world application. {Notably, maintaining identical construction time or memory consumption between baselines and \method for a fair comparison is impractical. Therefore, we provide further comparisons in the appendix.}

\subsubsection{\textbf{Parameter Sensitivity Study}}
\label{sec:exp-para-sen} \revision{Here, we examine how the performance of \method and the baselines change is affected by the two key hyperparameters, $M$ and $ef_{construction}$. 
Specifically, we increase $M$ and $ef_{construction}$ from 40, 200 to 50, 250, then to 60, 300 for \textsf{HNSW}$_{\textsf{F}}$, \textsf{HNSW}$_{\textsf{M}}$, and  \method. For \textsf{HNSW}$_{\textsf{O}}$, we adjust the $M$ and $ef_{construction}$ of each sub-index from 10, 50 to 12, 60, then to 14, 70, ensuring that they have comparative edges and allow for a fair comparison. 
Due to the page limitation, we only report the results for $\alpha \in [0, 0.2], [0.4, 0.6], [0.8, 1.0]$ when $M=40, 60$, with the other results in the appendix.} 
The results demonstrate that: \revision{(1) For the different number of edges $M$ and candidate set size $ef_{construction}$,  \method maintains the best performance among the baselines. (2) With varying $M$ and $ef_{construction}$, both baselines \textsf{HNSW}$_{\textsf{F}}$, \textsf{HNSW}$_{\textsf{M}}$, and  \method exhibit similar performance. This indicates that for graph-based ANNS indexes, once $M$ and $ef_{construction}$ are fine-tuned to relatively large values, the key factors determining performance are no longer the hyperparameters, but rather the edge selection strategy, as to be shown later. (3) The performance of the \textsf{HNSW}$_{\textsf{O}}$ improves as $M$ and $ef_{construction}$ increase, but but this leads to significantly higher index construction costs. For example, when $M = 14$ and $ef_{construction}=70$, the overall performance of \textsf{HNSW}$_{\textsf{O}}$ becomes comparable to that of \textsf{HNSW}$_{\textsf{F}}$, 
but the build time for \textsf{HNSW}$_{\textsf{O}}$ is 
twice as long as that of \textsf{HNSW}$_{\textsf{F}}$. This indicates that although \textsf{HNSW}$_{\textsf{F}}$ offers better performance as $M$ and $ef_{construction}$ increase, it comes at the cost of significantly higher indexing overhead. }



\begin{figure}[!t]
\centering
\subcaptionbox{\revision{$M=40, ef_{constrution}=200$ ($M=10, ef_{construction}=50$ for \textsf{HNSW}$_{\textsf{O}}$)}\label{fig:exp-openimage-40-200-full}}{
\includegraphics[width=0.7\textwidth]{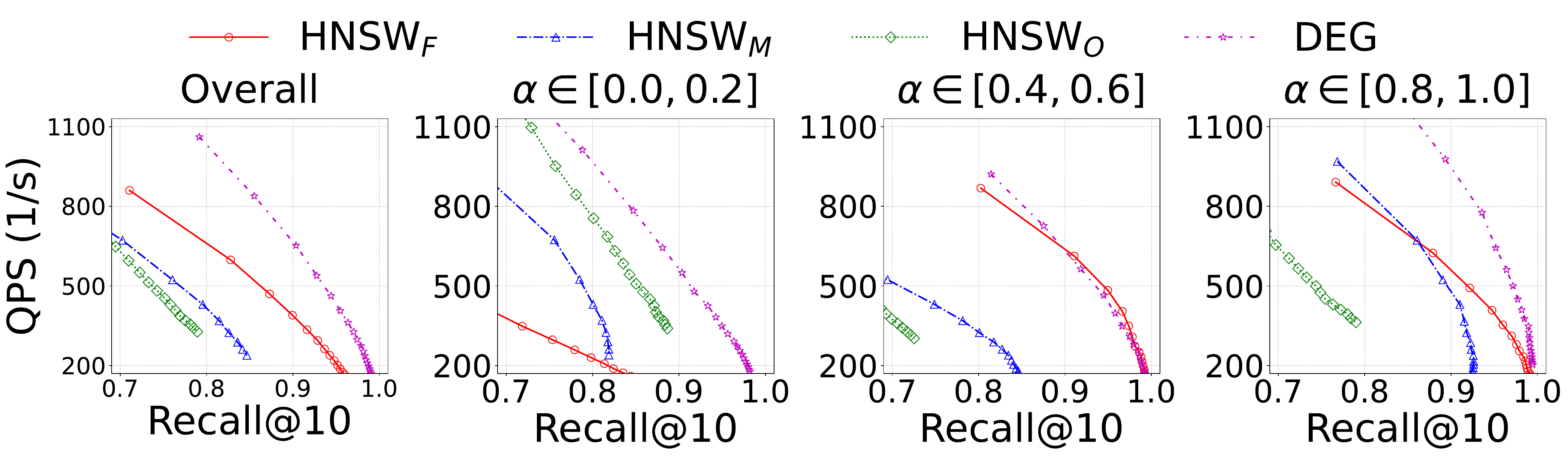}
}
\subcaptionbox{\revision{$M=60, ef_{constrution}=300$ ($M=14, ef_{construction}=70$ for \textsf{HNSW}$_{\textsf{O}}$)}\label{fig:exp-openimage-60-300-full}}{
\includegraphics[width=0.7\textwidth]{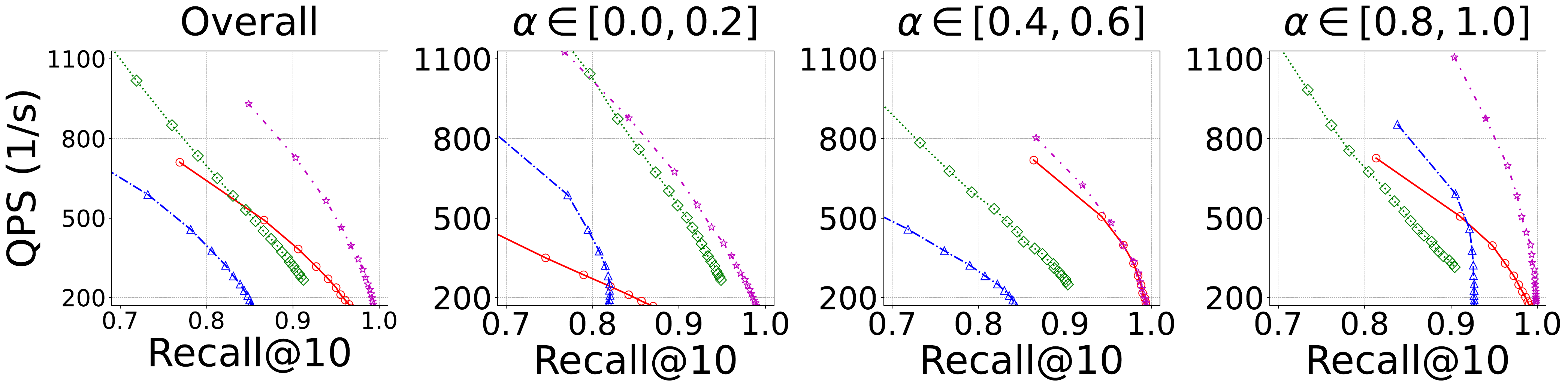}
}
\caption{\revision{The accuracy-efficiency trade-off results on \textsf{OpenImage} dataset with varying $M$ and $ef_{constrution}$.}}
\label{fig:exp-acc-eff-para-sensitivity-full}
\end{figure}

\subsubsection{\textbf{Ablation study}} 
\label{sec:exp-ablation-study} Here, we investigate how the proposed components contribute to \method's performance. Due to page limitations, we present results when \revision{$\alpha \in [0, 0.2], [0.4, 0.6], [0.8, 1]$}, with full results in the appendix.

\begin{figure}[!t]
\centering
\vspace*{-1em}
\includegraphics[width=0.7\textwidth]{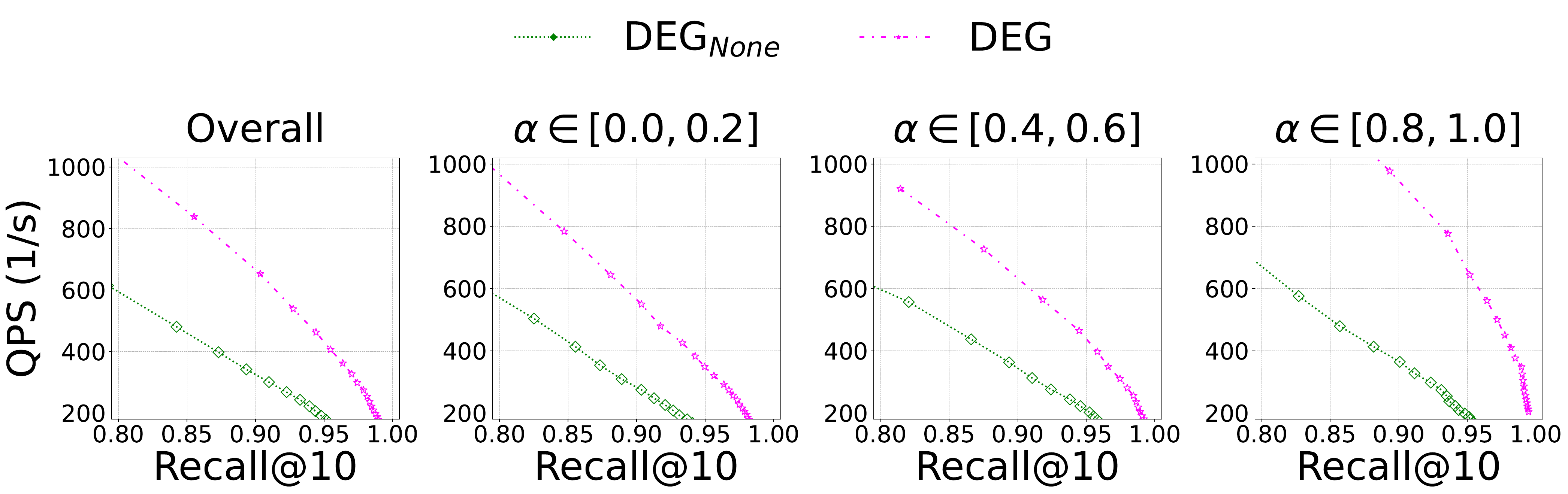}
\vspace*{-1em}
\caption{\revision{Ablation Study for the DEG's pruning strategy.}}
\vspace*{-1em}
\label{fig:exp-drng-ablation}
\end{figure}

\noindent\textbf{The \method's pruning strategy.}  To verify the effectiveness of the \method's pruning strategy, we compared \method with DEG$_{None}$, which does not apply any pruning strategy but uses the approximate 
Pareto frontiers obtained by the \GPS algorithm as edges directly and assigns each edge an active range $u=[0, 1]$. 
The experimental results are shown in Figure~\ref{fig:exp-drng-ablation}, which 
show that \method consistently outperforms DEG$_{None}$ by a large margin across varying $\alpha$. This validates the effectiveness of the \method's pruning strategy and confirms that the edge selection strategy is the key factor in enhancing the search performance of graph-based ANNS indexes.



\begin{figure}[!t]
\centering
\includegraphics[width=0.7\textwidth]{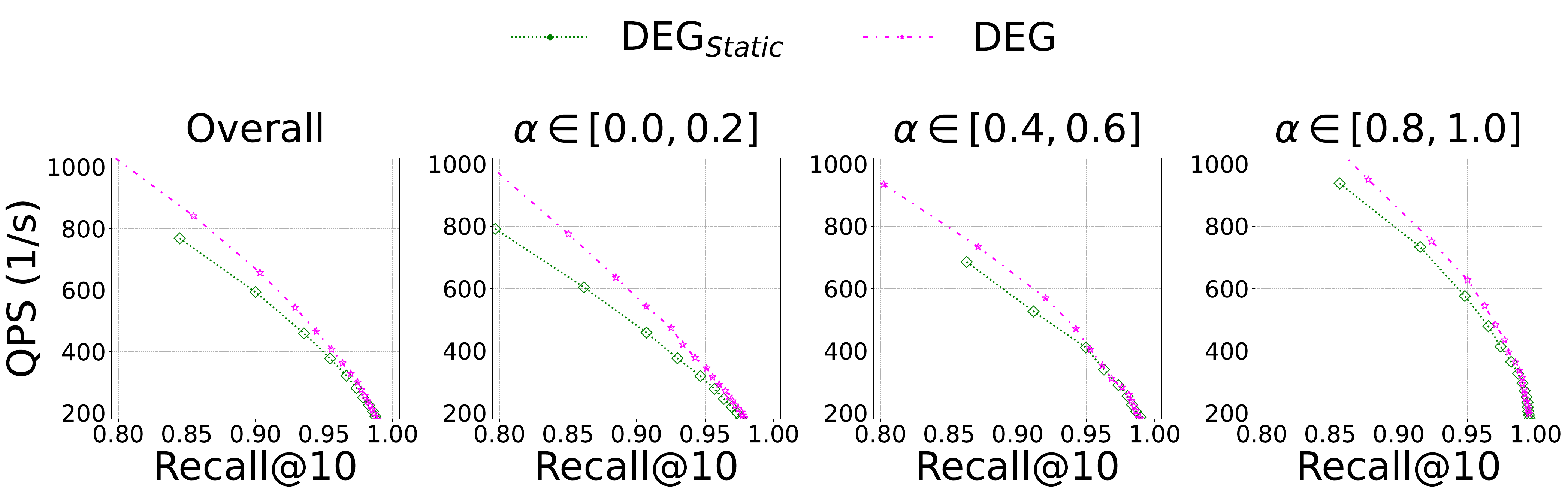}
\vspace*{-1em}
\caption{\revision{Ablation Study for the Active Range.}}
\vspace*{-1em}
\label{fig:exp-active-range-ablation}
\end{figure}

\noindent\textbf{The active range.} We further explore how the active range enhances \method's search performance by proposing an alternative method, DEG$_{static}$. DEG$_{static}$ uses the same index as \method but routes through all edges during the search phase, ignoring the active range. The experimental results on the \textsf{OpenImage} dataset are shown in Figure~\ref{fig:exp-active-range-ablation}. {The results show that the \method consistently outperforms the DEG$_{static}$ for varying $\alpha$, which indicates the active range can enhance search performance.}

\begin{figure}[!t]
\centering
\includegraphics[width=0.7\textwidth]{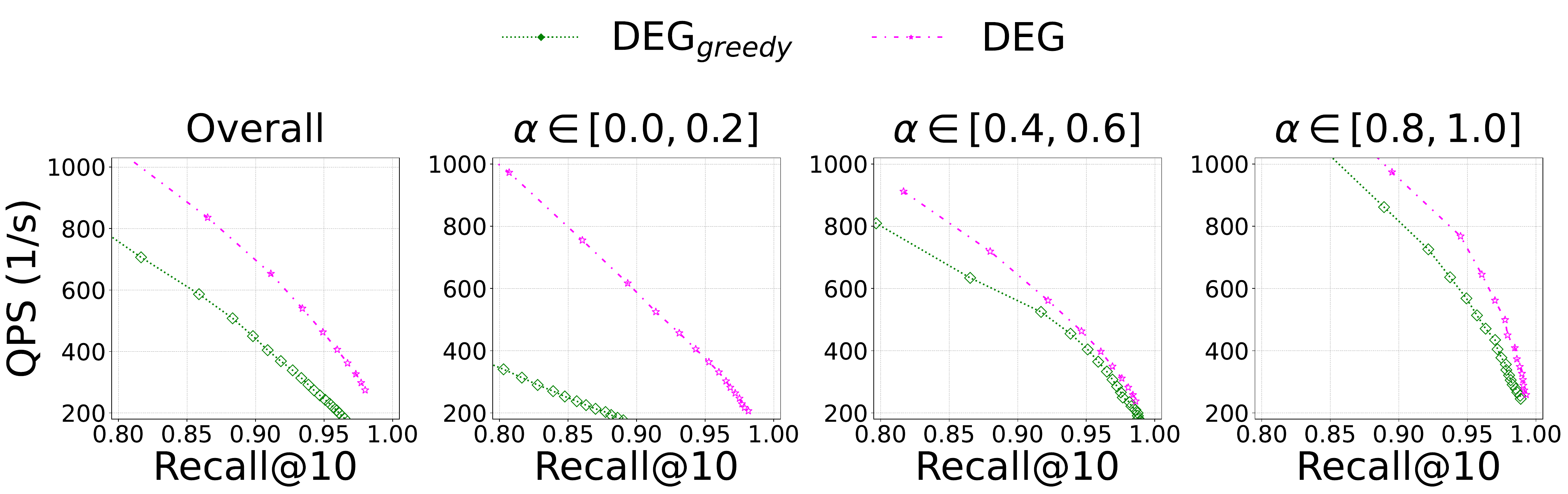}
\vspace*{-1em}
\caption{\revision{Ablation Study for candidate acquisition method.}}
\vspace*{-1em}
\label{fig:exp-gps-ablation}
\end{figure}

\noindent\textbf{The candidate acquisition method.}
To verify the effectiveness of the GPS algorithm, 
we replaced the GPS algorithm in \method with the greedy search algorithm, fixing $\alpha = 0.5$ during the index construction phase, which we call \textsf{DEG}$_{\textsf{greedy}}$. 
The experiment results on the \textsf{OpenImage} dataset are shown in Figure~\ref{fig:exp-gps-ablation}. {The results validate that \method consistently outperforms the \method$_{\textsf{greedy}}$ for varying $q.\alpha$, proving that the \GPS algorithm can acquire better candidates than the greedy search algorithm.}

\begin{figure}[!t]
\centering
\includegraphics[width=0.7\textwidth]{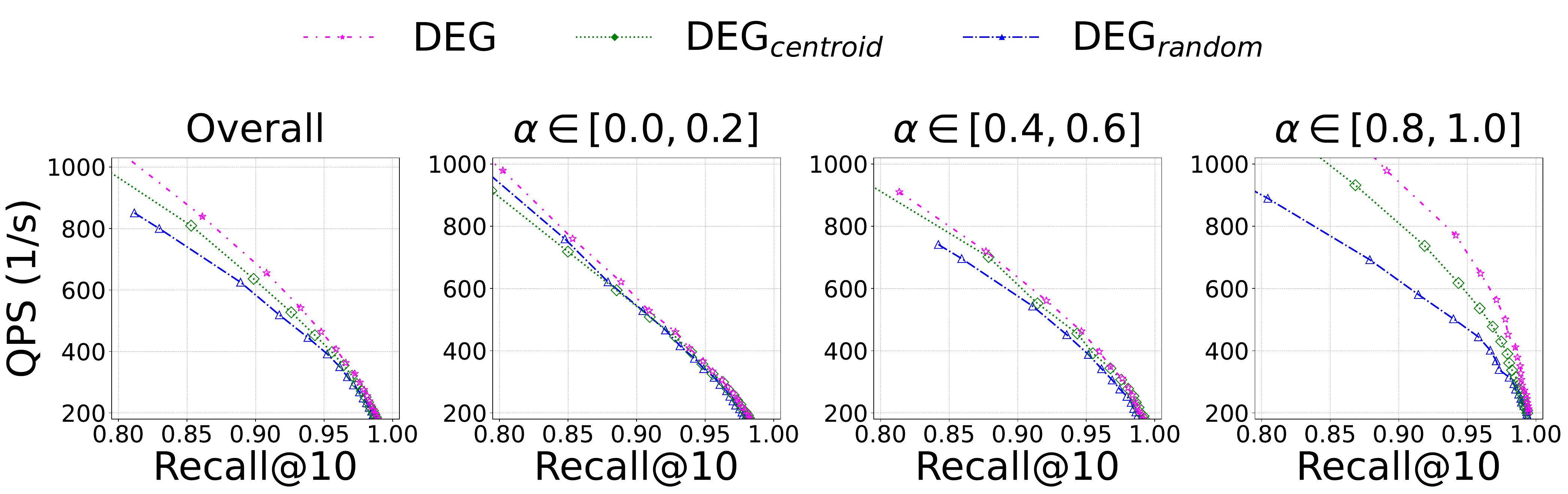}
\vspace{-1em}
\caption{\revision{Ablation Study for Edge Seed Acquisition method.}}
\vspace*{-1em}    
\label{fig:exp-seed-ablation}
\end{figure}

\noindent\textbf{The edge seed method.} To verify the effectiveness of the edge seed acquisition method, we 
consider two alternative methods, called DEG$_{\textsf{centroid}}$, which selects the group of points closest to the centroid as the seed. Specifically, it maintains the Pareto frontier of the graph center during the index construction phase and uses it as the starting point during the search phase. Another method is called DEG$_{\textsf{random}}$, which randomly selects some nodes, approximately the same size as the edge seed, as the starting point. The experimental results are shown in Figure~\ref{fig:exp-seed-ablation}. {The results show that \method consistently outperforms the two alternatives for varying $\alpha$, which validates the edge seed method's superiority.}

\section{CONCLUSIONS AND FUTURE WORK}
In this paper, we propose a novel ANNS index, \method, for the hybrid vector query problem, \revision{which comprises three novel components: the greedy Pareto frontier search algorithm, the dynamic edge pruning strategy, and the edge seed method.} 
One possible future direction is to extend our proposed index to multi-vector queries, where each object involves more than two vectors. Another potential direction is to develop a disk-based version of our index, making it suitable for implementation in vector databases.

\section*{Acknowledgments}
We thank the anonymous reviewers for their valuable feedback. This research is supported in part by Singapore MOE AcRF Tier-2 grants MOE-T2EP20221-0015 and MOE-T2EP20223-0004, and MOE AcRF Tier-1 grant RT6/23.

\newpage


\appendix

\newpage
\appendix
\revision{\section{Proof of Theorem 3.1}}

\begin{proof}
\label{theorem:nearest_proof} 
{For any $y \in D$ but $y \notin PF(D, p)$, there must exist $x \in PF(D, p)$ that dominates $y$; otherwise, $y$ would have been added to $PF(D, p)$ as well according to the definition. Consequently, we have $\delta_e(p, x) \leq \delta_e(p, y)$ and $\delta_s(p, x) \leq \delta_s(p, y)$, with at least one inequality being strict. Given that $\alpha \in [0,1]$ and the Euclidean distance is non-negative, we can derive that for any $\alpha$, $Dist(p, x) < Dist(p, y)$.}
\end{proof}

\revision{\section{Proof of Lemma 3.2}}
\begin{proof}
\label{lemma:drng_proof} 
{Since both Equation~\ref{eq:rng-hvq-1} and Equation~\ref{eq:rng-hvq-2} hold for any $\alpha \in r^z_1 \cap r^z_2$, then we can derive that for varying $\alpha \in r^z_1 \cap r^z_2$, the edge length of $(x, y)$, i.e., the distance between $x$ and $y$, is consistently larger than the edge lengths of both $(x, z)$ and $(y, z)$ as the right sides of the two inequalities represent the hybrid distance formulas for $(x, y)$, while the left sides represent the hybrid distance formulas for $(x, z)$ and $(y, z)$, respectively. Therefore, for any $\alpha \in r^z_1 \cap r^z_2$, the edge $(x, y)$ is consistently the longest edge in the triangle $(x, y, z)$, and will be pruned according to the RNG's pruning strategy.}
\end{proof}

\revision{\section{Proof of Lemma 3.3}}
\begin{proof}
\label{lemma:drng_nearest_neighbor_proof} 
{Given a specific $\alpha$, the graph formed by the active edges constructed using the above method will always be an exact RNG. This is because if any edge can be inserted into the graph without violating the RNG property for a specific $\alpha$ value, it would be included in the graph with an active range that includes $\alpha$ according to our dynamic edge pruning strategy. This ensures that no edges are omitted, thus guaranteeing that the constructed graph is an exact RNG. According to~\cite{fu2019fast, wang2021comprehensive}, exact RNG is equivalent to exact MRNG, thereby guaranteeing that the nearest neighbor can always be found for any query using the greedy search algorithm.}

\end{proof}

\section{Finding $l$-layer Pareto frontiers}
\label{sec:appendix-l-skyline}
\begin{algorithm}[t]
    \caption{Finding Pareto Frontiers}
    \small
    \label{alg:find-pf}
    \KwIn{A candidate pool $CS$ for node $p$, candidate pool size $ef_{construction}$}    
    \KwOut{$l$ layer Pareto frontiers $\{PF^{1}(p), \cdots, PF^{l}(p)\}$ for node $p$, where $\sum_{i = 1}^l |PF^{i}(p)| \leq ef_{construction}$}
    \BlankLine
    
    \SetKwFunction{FindPF}{\textsc{FindPF}}
    \SetKwProg{Fn}{Procedure}{:}{}
    \Fn{\FindPF{$CS, ef_{construction}$}}{
        \text{Sort} $x \in CS$ \text{in ascending order of} $\delta_s(x, p)$ \text{to} $p$\;
        $Res \leftarrow \emptyset$ \tcp*{\textsf{result set}}
        \While{$|Res| < ef_{construction}$}{
            $PF \leftarrow \emptyset$ \tcp*{\textsf{current pareto frontier}}
            $Remain \leftarrow \emptyset$ \tcp*{\textsf{remaining candidates}}
            $PrevEmbDist \leftarrow \infty$ \tcp*{\textsf{smallest $\delta_e(x, p)$, $x \in PF$}}   
            \ForEach{$x \in CS$} {
                \If{$\delta_e(x, p) < \text{PrevEmbDist}$} {
                    $PF.add(x)$\;
                    $PrevEmbDist \leftarrow \delta_e(x, p)$\;
                }
                \Else {
                    $Remain.add(x)$\;
                }
            }
            \If{$|PF| + |Res| < ef_{construction}$} {
                $Res.add(PF)$\;
                $CS \leftarrow Remain$\;
            }
            \Else {
                \textbf{break}\;
            }        
        }
        \Return $Res$ \;
    }
\end{algorithm}

Algorithm~\ref{alg:find-pf} summarizes the procedure of finding $l$-layer $PF(D, p)$ using the existing algorithm~\cite{borzsony2001skyline}. Initially, the entire set is sorted in ascending order of $\delta_s(x, p)$, and the result set $Res$ is initialized (lines 2-3). Then, it repeatedly finds the Pareto frontier and removes it from the candidate set until the result set reaches the size bound (lines 4-18). 
Specifically, in each iteration, two empty sets, $PF$, and $Remain$, are initialized to store the Pareto frontier and remaining nodes, respectively. A variable $PrevEmbDist$ is also initialized to store the $\delta_e(p, x)$ of the last added element in $PF$ (lines 5-7). The candidate set $CS$ is scanned to check if $\delta_e(p, x)$ is smaller than $PrevEmbDist$, determining if it should be added to $PF$ (lines 8-9). If added, $PrevEmbDist$ is updated; otherwise, the element is added to $Remain$ (lines 10-13). This process, proven correct in~\cite{borzsony2001skyline}, ensures that the next element added to $PF$ will have a smaller $\delta_e(p, x)$ value; otherwise, it is dominated by the previous element in $PF$ due to the sorted order. 
Finally, we check if the result set size exceeds the candidate set size bound $ef_{construction}$ (line 14). If it does, we break the loop and return $Res$ (lines 17-18). Otherwise, we add $PF$ to $Res$, update the candidate set $CS$ with $Remain$ (lines 15-16), and continue finding Pareto frontiers within $Remain$. The time complexity of Algorithm~\ref{alg:find-pf} is $O(|CS|\log(|CS|)+l|CS|)$, where $l$ is a constant. 
The proof is straightforward and therefore omitted.

\begin{figure}[!htbp]
\centering
\includegraphics[width=0.45\textwidth]{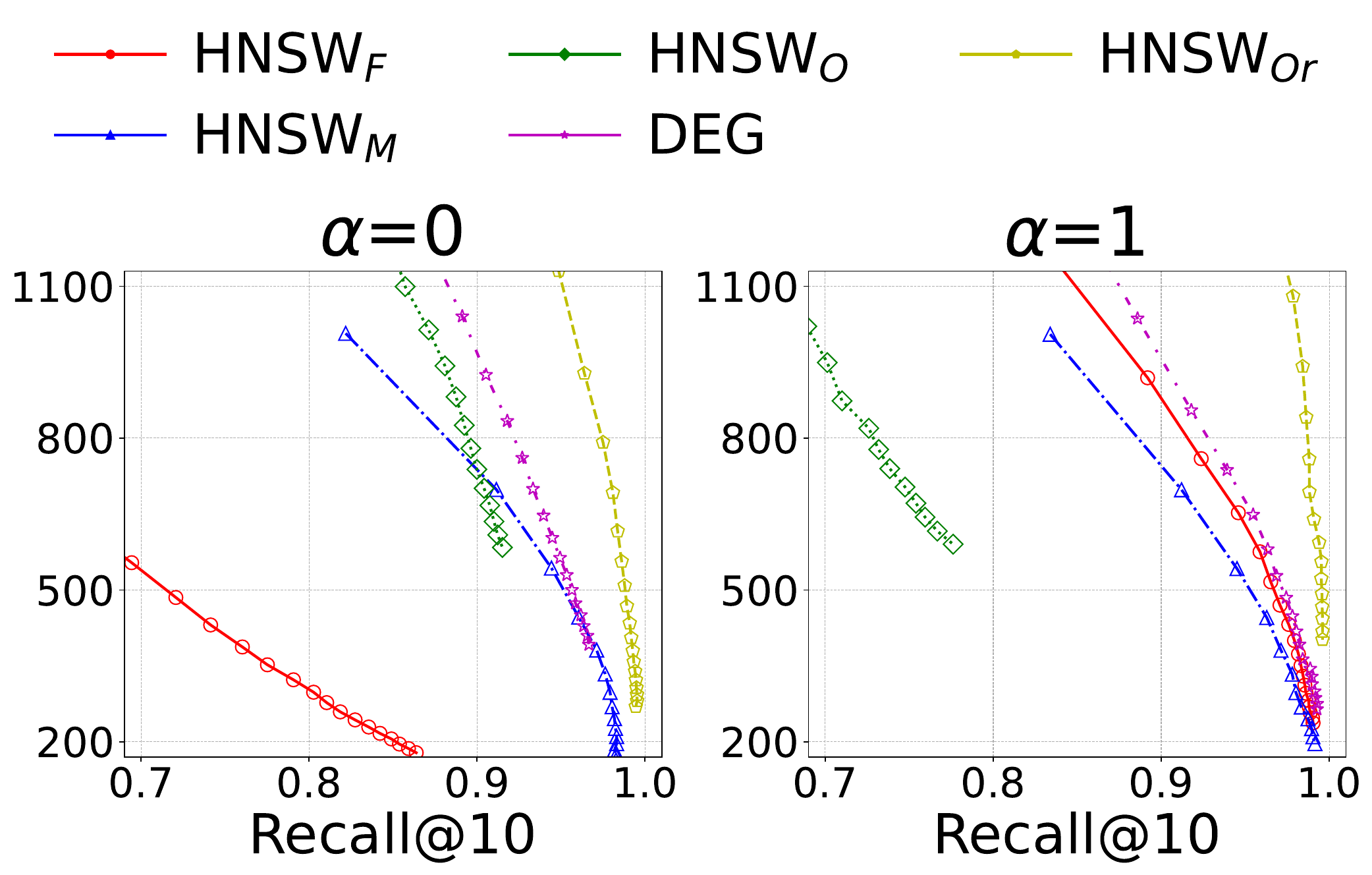}
\caption{\revision{The accuracy-efficiency trade-off results for the extreme query.}}
\label{fig:exp-01}
\end{figure}

\revision{\section{Normal Vector Query Study}}
\noindent\revision{Here, we present the experimental results for the normal vector query case, where $q.\alpha = 0$ and $1$. It is worth noting that such a case is not a common scenario for \hvq. For example, \cite{liu2023} finds that the optimal weight for most queries ranges from 0.1 to 0.4. Given that our proposed \method is tailored for \hvq, it is as expected that \method performs worse than the ideal method, \textsf{oracle}, which constructs separate HNSW indexes for each modality and conducts searches within the corresponding index (as detailed in Section 5.1). Our expectation is that it should outperform the existing baselines for \hvq. 
The experimental results in Figure~\ref{fig:exp-01}
show that \method
outperforms the existing baselines of \hvq. 
The results demonstrate that our proposed DEG achieves the best performance compared to the baselines. 
Specifically, \textsf{HNSW}$_\textsf{F}$ shows competitive performance at $\alpha = 1$ but declines significantly when $\alpha = 0$. A similar pattern is observed for \textsf{HNSW}$_\textsf{O}$, aligning with our observation in Section 4.2. The explanation can be found in Section 4.2.1, observation (2). \textsf{HNSW}$_\textsf{M}$ performs well at both $\alpha = 0$ and $\alpha = 1$ because its two indexes are constructed specifically under the two values. DEG consistently outperforms the performance of \textsf{HNSW}$_\textsf{M}$, demonstrating its capability to handle normal vector queries. 
As expected, the ideal method \textsf{HNSW}$_\textsf{Or}$ outperforms \method, along with all other baselines.}

\section{Sensitivity Study}
The complete experimental results of the sensitivity study are presented in Figure~\ref{fig:exp-acc-eff-para-sensitivity-full}, confirming our previous statement.

\begin{table}[!htbp]
    \centering
    \caption{\ready{The Indexing Time for the \textsf{OpenImage} Dataset (Min).}}
    \label{tab:index-time}
    \renewcommand{\arraystretch}{1.2} 
    \setlength{\arrayrulewidth}{0.5pt}
    \begin{tabularx}{1.0\textwidth}{>{\centering}m{8cm}*{4}{X}}
    \toprule[1pt]
    $M, ef_{constrution}$ & \textsf{HNSW}$_{\textsf{F}}$ & \textsf{HNSW}$_{\textsf{M}}$ & \textsf{HNSW}$_{\textsf{O}}$ & \method\\
    \midrule[1pt]
    $40, 200$ for \textsf{HNSW}$_{\textsf{F}}$, \textsf{HNSW}$_{\textsf{M}}$ and $10, 50$ for \textsf{HNSW}$_{\textsf{O}}$ & 40 & 45 & 47 & 75 \\
    $50, 250$ for \textsf{HNSW}$_{\textsf{F}}$, \textsf{HNSW}$_{\textsf{M}}$ and $12, 60$ for \textsf{HNSW}$_{\textsf{O}}$ & 46 & 51 & 58 & N/A 
    \\
    $60, 300$ for \textsf{HNSW}$_{\textsf{F}}$, \textsf{HNSW}$_{\textsf{M}}$ and $14, 70$ for \textsf{HNSW}$_{\textsf{O}}$ & 52 & 58 & 70 & N/A 
    \\ 
    $70, 350$ ffor \textsf{HNSW}$_{\textsf{F}}$, \textsf{HNSW}$_{\textsf{M}}$ and $16, 80$ for \textsf{HNSW}$_{\textsf{O}}$ & 58 & 62 & 78 & N/A \\ 
    $80, 400$ for \textsf{HNSW}$_{\textsf{F}}$, \textsf{HNSW}$_{\textsf{M}}$ and $18, 90$ for \textsf{HNSW}$_{\textsf{O}}$ & 62 & 66 & 82 & N/A \\ 
    $20, 100$ for \textsf{HNSW}$_{\textsf{O}}$ & N/A & N/A & 90 & N/A \\ 
    $30, 150$ for \textsf{HNSW}$_{\textsf{O}}$ & N/A & N/A & 120 & N/A \\ 
    $40, 200$ for \textsf{HNSW}$_{\textsf{O}}$ & N/A & N/A & 176 & N/A \\ 
    \bottomrule[1pt]
  \end{tabularx}
\end{table}

\begin{table}[!htbp]
    \centering
    \caption{\ready{The Memory Consumption for the \textsf{OpenImage} Dataset (MB).}}
    \label{tab:index-memory}
    \renewcommand{\arraystretch}{1.2} 
    \setlength{\arrayrulewidth}{0.5pt}
    \begin{tabularx}{1.0\textwidth}{>{\centering}m{8cm}*{4}{X}}
    \toprule[1pt]
    $M, ef_{constrution}$ & \textsf{HNSW}$_{\textsf{F}}$ & \textsf{HNSW}$_{\textsf{M}}$ & \textsf{HNSW}$_{\textsf{O}}$ & \method\\
    \midrule[1pt]
    $40, 200$ for \textsf{HNSW}$_{\textsf{F}}$, \textsf{HNSW}$_{\textsf{M}}$ and $10, 50$ for \textsf{HNSW}$_{\textsf{O}}$ & 3220 & 3412 & 3614 & 4203 \\
    $50, 250$ for \textsf{HNSW}$_{\textsf{F}}$, \textsf{HNSW}$_{\textsf{M}}$ and $12, 60$ for \textsf{HNSW}$_{\textsf{O}}$ & 3238 & 3418 & 3639 & N/A 
    \\
    $60, 300$ for \textsf{HNSW}$_{\textsf{F}}$, \textsf{HNSW}$_{\textsf{M}}$ and $14, 70$ for \textsf{HNSW}$_{\textsf{O}}$ & 3253 & 3420 & 3654 & N/A 
    \\ 
    $70, 350$ for \textsf{HNSW}$_{\textsf{F}}$, \textsf{HNSW}$_{\textsf{M}}$ and $16, 80$ for \textsf{HNSW}$_{\textsf{O}}$  & 3271 & 3443 & 3663 & N/A \\ 
    $80, 400$ for \textsf{HNSW}$_{\textsf{F}}$, \textsf{HNSW}$_{\textsf{M}}$ and $18, 90$ for \textsf{HNSW}$_{\textsf{O}}$ & N/A
    & N/A
    & 3759 & N/A \\ 
    $20, 100$ for \textsf{HNSW}$_{\textsf{O}}$ & N/A & N/A & 3793 & N/A \\ 
    $30, 150$ for \textsf{HNSW}$_{\textsf{O}}$ & N/A & N/A & 3879 & N/A \\ 
    $40, 200$ for \textsf{HNSW}$_{\textsf{O}}$ & N/A & N/A & 4080 & N/A \\
    \bottomrule[1pt]
  \end{tabularx}
\end{table}

\begin{table}[!htbp]
    \centering
    \caption{\ready{Average number of edges per node for each index in the \textsf{OpenImage} dataset.}}
    \label{tab:index-edge-number}
    \renewcommand{\arraystretch}{1.2} 
    \setlength{\arrayrulewidth}{0.5pt}
    \begin{tabularx}{1.0\textwidth}{>{\centering}m{8cm}*{4}{X}}
    \toprule[1pt]
    $M, ef_{constrution}$ & \textsf{HNSW}$_{\textsf{F}}$ & \textsf{HNSW}$_{\textsf{M}}$ & \textsf{HNSW}$_{\textsf{O}}$ & \method\\
    \midrule[1pt]
    $40, 200$ for \textsf{HNSW}$_{\textsf{F}}$, \textsf{HNSW}$_{\textsf{M}}$ and $10, 50$ for \textsf{HNSW}$_{\textsf{O}}$ & 24 & 42 & 35 & 37 \\
    $50, 250$ for \textsf{HNSW}$_{\textsf{F}}$, \textsf{HNSW}$_{\textsf{M}}$ and $12, 60$ for \textsf{HNSW}$_{\textsf{O}}$ & 31 & 47 & 44 & N/A 
    \\
    $60, 300$ for \textsf{HNSW}$_{\textsf{F}}$, \textsf{HNSW}$_{\textsf{M}}$ and $14, 70$ for \textsf{HNSW}$_{\textsf{O}}$ & 32 & 51 & 51 & N/A 
    \\ 
    $70, 350$ for \textsf{HNSW}$_{\textsf{F}}$, \textsf{HNSW}$_{\textsf{M}}$ and $16, 80$ for \textsf{HNSW}$_{\textsf{O}}$  & 32 & 51 & 63 & N/A \\ 
    $80, 400$ for \textsf{HNSW}$_{\textsf{F}}$, \textsf{HNSW}$_{\textsf{M}}$ and $18, 90$ for \textsf{HNSW}$_{\textsf{O}}$ & N/A & N/A & 72 & N/A \\     
    $20, 100$ for \textsf{HNSW}$_{\textsf{O}}$ & N/A & N/A & 84 & N/A \\ 
    $30, 150$ for \textsf{HNSW}$_{\textsf{O}}$ & N/A & N/A & 95 & N/A \\ 
    $40, 200$ for \textsf{HNSW}$_{\textsf{O}}$ & N/A & N/A & 117 & N/A \\ 
    \bottomrule[1pt]
  \end{tabularx}
\end{table}

\section{{Different Hyperparameter Study}}
\ready{In Table~\ref{tab:index-time}, Table~\ref{tab:index-memory}, and Table~\ref{tab:index-edge-number}, we report the construction time, memory usage, and average number of edges per node for each baseline index under various hyperparameter settings on the \textsf{OpenImage} dataset, and compare these results with those of our proposed \method configured with $M = 40$ and $ef_{constrution} = 200$. Notably, we optimize the implementation of our proposed \method by using integers instead of floats to store the active range for each edge, thereby reducing memory consumption. We have the following findings: (1) For the baselines \textsf{HNSW}$_{\textsf{F}}$ and \textsf{HNSW}$_{\textsf{M}}$, increasing $M$ and $ef_{construction}$ does not lead to a significant increase in memory consumption. This is because the memory usage is primarily dominated by the high-dimensional vectors, and the average number of edges per node for these indexes remains relatively stable. We do not increase the hyperparameters $M$ and $ef_{construction}$ further, as search performance does not show any additional improvement. We attribute this phenomenon to the lack of increase in the average number of edges.
(2) For baseline \textsf{HNSW}$_{\textsf{O}}$, as $M$ and $ef_{construction}$ increase, the construction time and memory consumption of this baseline increase significantly. 
We increase the hyperparameters of \textsf{HNSW}$_\textsf{O}$ to $M = 40$ and $ef_{construction} = 200$, but do not raise them further, as its memory consumption is comparable to our method, while construction time has more than doubled compared to our method. }

\begin{figure*}[!htbp]
\centering
\subcaptionbox{\revision{$M=40, ef_{constrution}=200$ for \method, $M=80, ef_{construction}=400$ for \textsf{HNSW}$_{\textsf{F}}$ and \textsf{HNSW}$_{\textsf{M}}$, and $M=30, ef_{construction}=150$ for \textsf{HNSW}$_{\textsf{O}}$}\label{fig:exp-openimage-40-200-full}}{
\includegraphics[width=1.0\textwidth]{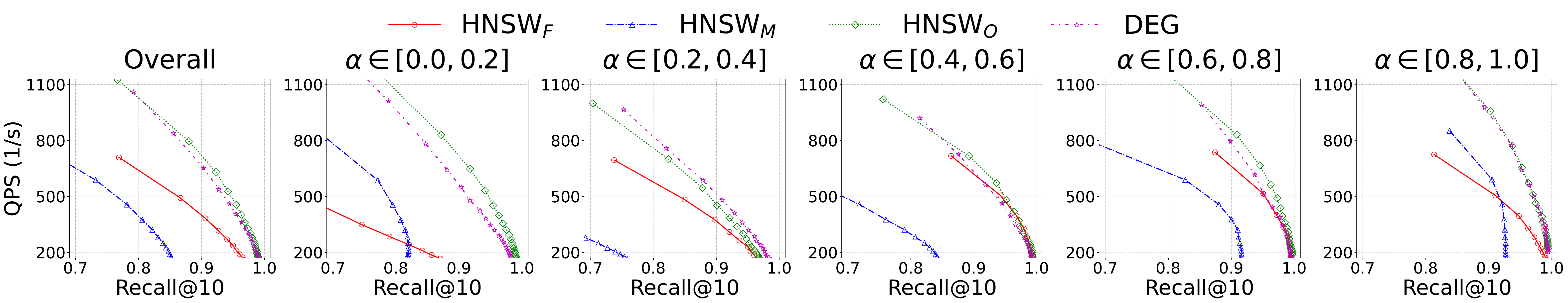}
}
\caption{\ready{The accuracy-efficiency trade-off results on \textsf{OpenImage} dataset with different $M$ and $ef_{constrution}$.}}
\label{fig:exp-acc-eff-para-different-full}
\end{figure*}

\ready{In Figure~\ref{fig:exp-acc-eff-para-different-full}, we compare our proposed \method, configured with $M = 40$ and $ef_{construction} = 200$, to \textsf{HNSW}$_\textsf{M}$ and \textsf{HNSW}$_\textsf{F}$, both set to $M = 70$ and $ef_{construction} = 350$, as well as \textsf{HNSW}$_\textsf{O}$, configured with $M = 40$ and $ef_{construction} = 200$. 
The results demonstrate that our proposed \method consistently achieves superior performance under different hyperparameters. Specifically, our proposed \method consistently outperforms \textsf{HNSW}$_\textsf{M}$ and \textsf{HNSW}$_\textsf{F}$. Compared to \textsf{HNSW}$_\textsf{O}$, our proposed \method achieves similar performance while requiring only half the time for index construction.}

\begin{figure*}[!htbp]
\centering
\subcaptionbox{\revision{$M=40, ef_{constrution}=200$ ($M=10, ef_{construction}=50$ for \textsf{HNSW}$_{\textsf{O}}$)}\label{fig:exp-openimage-40-200-full}}{
\includegraphics[width=1.0\textwidth]{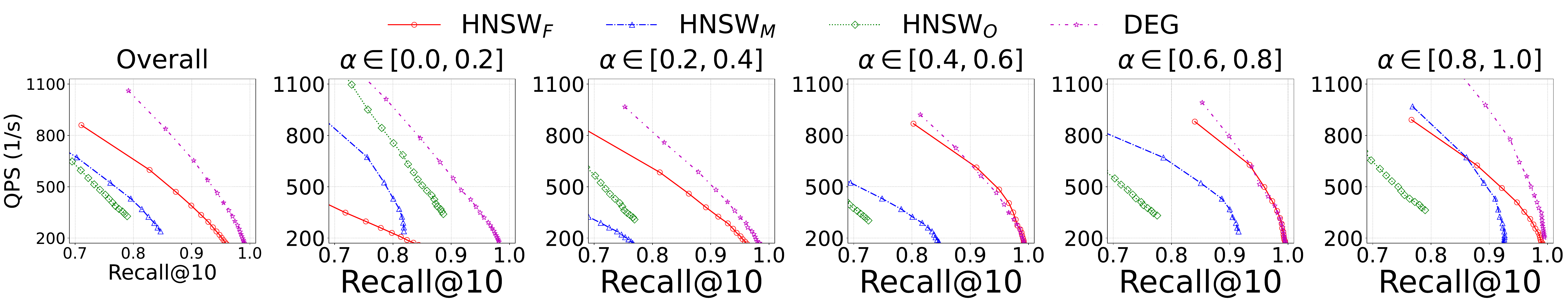}
}
\subcaptionbox{\revision{$M=50, ef_{constrution}=250$ ($M=12, ef_{construction}=60$ for \textsf{HNSW}$_{\textsf{O}}$)\label{fig:exp-openimage-50-250-full}}}{
\includegraphics[width=1.0\textwidth]{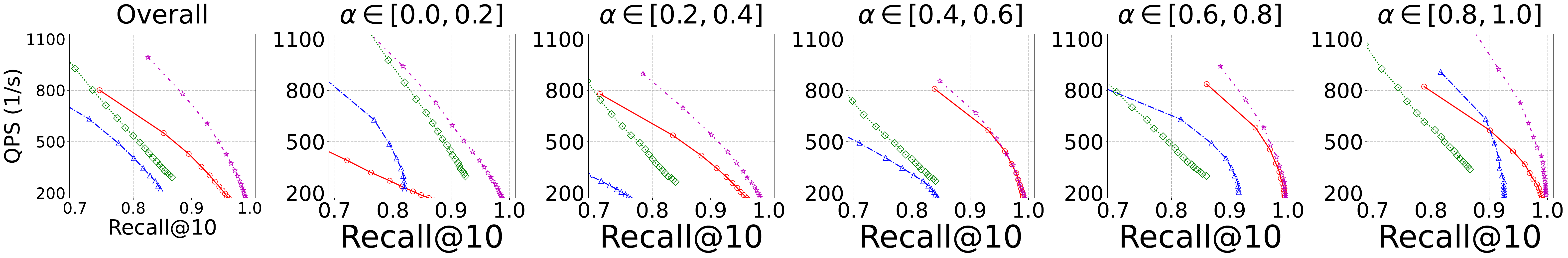}
}
\subcaptionbox{\revision{$M=60, ef_{constrution}=300$ ($M=14, ef_{construction}=70$ for \textsf{HNSW}$_{\textsf{O}}$)}\label{fig:exp-openimage-60-300-full}}{
\includegraphics[width=1.0\textwidth]{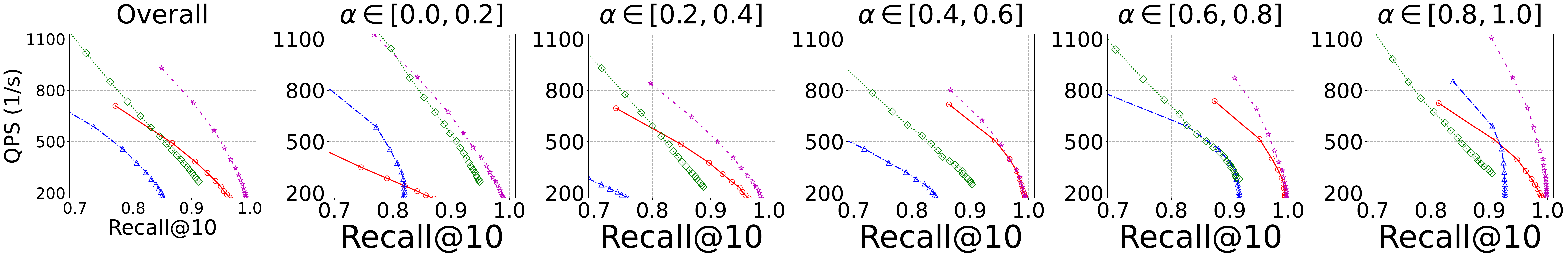}
}
\caption{\revision{The accuracy-efficiency trade-off results on \textsf{OpenImage} dataset with varying $M$ and $ef_{constrution}$.}}
\label{fig:exp-acc-eff-para-sensitivity-full}
\end{figure*}

\begin{figure*}[!htbp]
\centering
\includegraphics[width=1.0\textwidth]{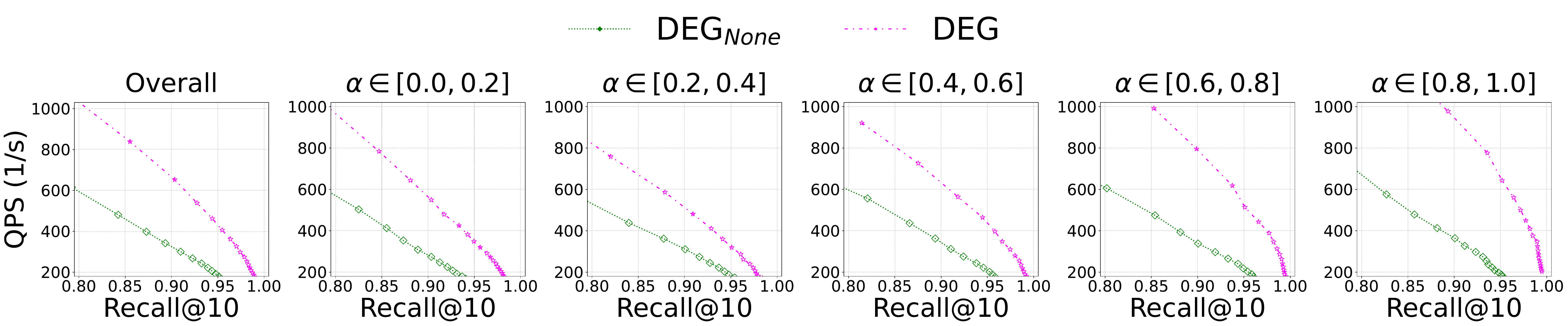}
\caption{\revision{Ablation Study for the D-RNG's pruning strategy on the \textsf{OpenImage} dataset.}}
\label{fig:exp-drng-ablation-full}
\end{figure*}

\begin{figure*}[!htbp]
\centering
\includegraphics[width=1.0\textwidth]{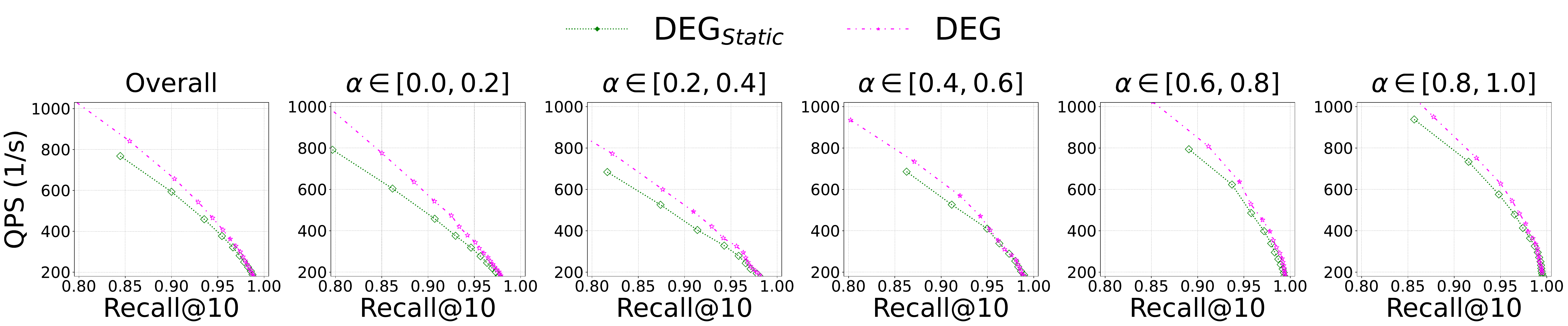}
\caption{\revision{Ablation Study for the Active Range on the \textsf{OpenImage} dataset.}}
\label{fig:exp-active-range-ablation-full}
\end{figure*}

\begin{figure*}[!htbp]
\centering
\includegraphics[width=1.0\textwidth]{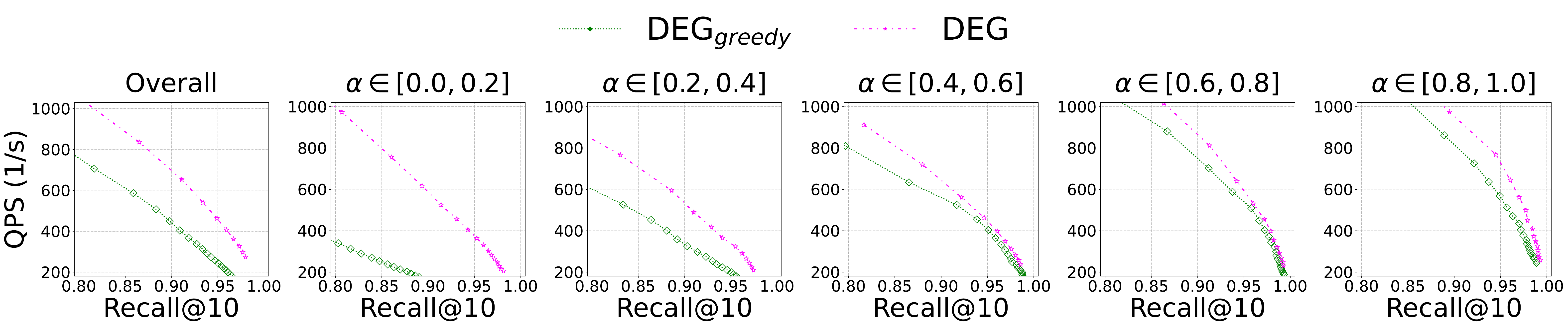}
\caption{\revision{Ablation Study for the candidate acquisition method on the \textsf{OpenImage} dataset.}}
\label{fig:exp-gps-ablation-full}
\end{figure*}

\begin{figure*}[!htbp]
\centering
\includegraphics[width=1.0\textwidth]{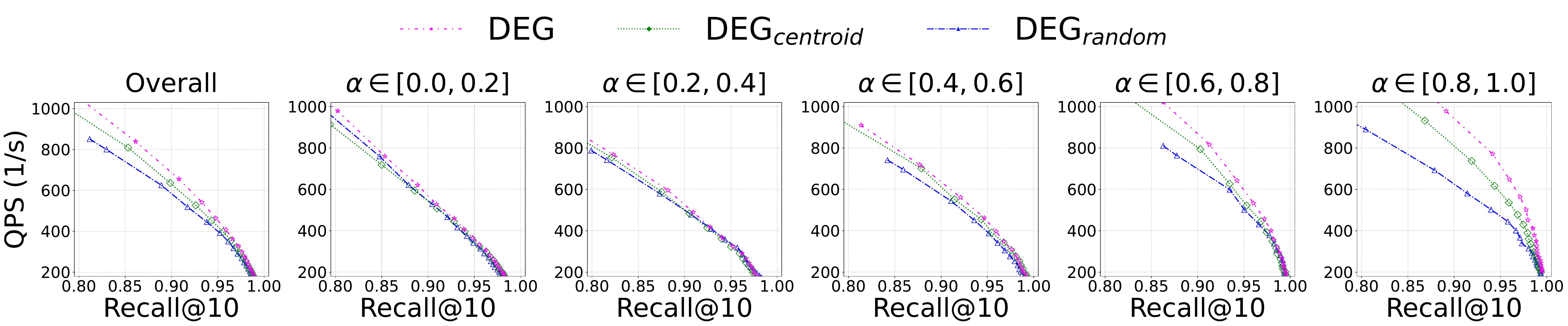}
\caption{\revision{Ablation Study for the Edge Seed Acquisition method on the \textsf{OpenImage} dataset.}}
\label{fig:exp-seed-ablation-full}
\end{figure*}

\section{Ablation Study}
The complete experimental results of the ablation study are presented from Figure~\ref{fig:exp-drng-ablation-full} to Figure~\ref{fig:exp-seed-ablation-full}, which is consistent with our previous statement.



\section{Semantic-aware Spatial keyword query}
\label{sec:appendix-stkq}
Recently, some studies~\cite{chenS2RtreePivotbasedIndexing2020, qianSemanticawareTopkSpatial2018, DBLP:conf/edbt/TheodoropoulosN24} have attempted to address the special case of the \hvq problem when one of the feature vectors has a small dimensionality (e.g., m = 2 or 3). However, these methods all suffer from significant efficiency issues. Details of these methods are discussed below.

\begin{itemize}[leftmargin=*]
    \item \textsf{NIQ~\cite{qianSemanticawareTopkSpatial2018}}: This method employs a Quadtree~\cite{finkelQuadTreesData1974} to index low-dimensional feature vectors ($o.s$) and an iDistance index~\cite{Jagadishidistance2005} to organize high-dimensional feature vectors ($o.e$) within each Quadtree leaf node. During the querying phase, it first traverses the Quadtree to find nearby leaf nodes and then uses the iDistance index within the leaf nodes to find exact results.
    \item \textsf{S2R~\cite{chenS2RtreePivotbasedIndexing2020}}: This method uses an R-tree~\cite{beckmann1990r} to index low-dimensional feature vectors and a pivot-based mechanism to map high-dimensional feature vectors to lower-dimensional vectors, which are then organized by another R-tree. During the querying phase, similar to the NIQ-Tree, it hierarchically traverses R-trees to locate exact results.
    \item \textsf{CSSI~\cite{DBLP:conf/edbt/TheodoropoulosN24}}: This method uses PCA to project high-dimensional feature vectors, clusters the low-dimensional and PCA-projected vectors separately with K-means, and forms hybrid clusters by combining the resulting clusters. The centroids and radii of the clusters are computed from the mean of the objects' feature vectors. During the querying phase, it routes and prunes clusters based on the distance between the query and cluster centroids, and produces exact results by reranking the visited objects.
    \item \textsf{CSSIA~\cite{DBLP:conf/edbt/TheodoropoulosN24}}: This method modifies CSSI by using the mean of PCA vectors for cluster centroids, while still using original vectors for reranking within clusters, thereby producing approximate results. It stops the search only when it believes no better candidates remain based on the distance to the PCA centroids, thereby lacking the accuracy-efficiency trade-off.
\end{itemize}
However, none of these methods can handle cases where both vectors are high-dimensional. Furthermore, our experiments show that these methods have time costs similar to those of brute-force search, making them unsuitable for retrieval tasks. 

\end{document}